\documentclass[12pt]{article}
\makeatletter
\@addtoreset{equation}{section}
\makeatother

\newcommand{\be}{\begin{equation}}
\newcommand{\ee}{\end{equation}}
\newcommand{\bea}{\begin{eqnarray}}
\newcommand{\eea}{\end{eqnarray}}

\def\bb{\bar{\beta}}

\def\gt{\tilde{g}}

\def\Ph{\varphi}
\def\Pht{\tilde{\varphi}}
\renewcommand{\t}[1]{\tilde{#1}}

\begin{document}

\begin{titlepage}

\begin{center}
\vskip 2.5 cm
{\Large \bf Poisson-Lie Duality in the String Effective Action }
\vskip 1 cm
{\large A. Bossard\footnote{e-mail:\ \tt bossard@celfi.phys.univ-tours.fr} 
and N. Mohammedi\footnote{e-mail:\ \tt nouri@celfi.phys.univ-tours.fr}}\\
\vskip 1cm
{\em Laboratoire de Math\'ematiques et Physique Th\'eorique\footnote{CNRS UMR 6083}\\
Universit\'e Francois Rabelais \\
Facult\'e des Sciences et Techniques \\
Parc de Grandmont \\
F-37200 Tours - France.} 
\end{center}
\vskip 1.5 cm

\begin{abstract}  
The symmetry properties of the bosonic string effective action under Poisson-Lie
duality transformations are investigated. A convenient and simple formulation of 
these duality transformations is found, that allows the reduction of the string effective 
action in a Kaluza-Klein framework. It is shown that the action is invariant provided that 
the two Lie algebras, forming the Drinfeld double, have traceless structure constants.
Finally, a functional relation is found between the Weyl anomaly coefficients of
the original and dual non-linear sigma models. 
\end{abstract}
\end{titlepage}

\section{Introduction}

Duality transformations have played an important role in modern string theories.
Target space duality (T-duality), in particular, connects seemingly different backgrounds 
in which the strings can propagate \cite{buscher}. String backgrounds and their 
dual fields can be considered as different descriptions of the same physical system. In other words, 
they represent the  same point in moduli space of a given string theory.
It is therefore crucial to analyse the different forms of T-duality in order to better understand the 
moduli space features of string theory.

The T-duality transformations (unlike S-duality) are formulated at the level 
of the two-dimensional non-linear sigma model. 
There are, however, no general methods for constructing 
these transformations. 
One of the leading organizing principle in the first studies of T-duality is 
undoubtedly the notion of symmetries. 
If the background fields of the sigma model possess Abelian isometries
then it has been shown \cite{Rocek:1992ps} 
that obtaining of the dual theory is straightforward:
One gauges these symmetries and imposes zero curvature constraints on the gauge fields 
by means of Lagrange multipliers.  
The elimination of the gauge fields through their equations of motion, 
after a convenient gauge fixing, leads to the dual sigma model. 
The Lagrange multipliers are promoted to propagating fields in this dual theory.
An important feature of the Abelian case is the possibility to reverse this procedure 
and obtain back the original model from the dual one. 
This is due to the fact that Abelian T-duality preserves the symmetries of the original 
sigma model. For sigma models with background fields possessing 
non-Abelian isometries an exact analogue of the above 
method has been shown to apply as well \cite{delaOssa:1993vc}. 
However, the symmetries of the original theory are not preserved and
non-Abelian T-duality is not reversible \cite{Giveon:1994ai}.  
The formulation of these two kind of T-dualities in terms of canonical 
transformations \cite{Alvarez:1994wj,Curtright:1994be} is a check on 
the classical equivalence of a sigma model and its dual.

A major step in the search for criteria to find duality 
transformations connecting non-linear sigma models was 
made by Klimcik and Severa \cite{Klimcik}. It deals with sigma 
models based on two Lie groups and the corresponding duality transformations
are known as Poisson-Lie T-duality. Their method 
relies no longer on the presence of symmetries in the backgrounds
of the sigma model. It uses, instead, a very specific feature 
of duality: Namely, the interchange between the equations of motion 
of the original theory and the Bianchi identities of the dual model. 
They discovered that two non-linear sigma models, built on 
two Lie groups, are dual to each other when the two 
corresponding Lie algebras form a Drinfeld double. 
A further step towards the understanding of the quantum 
aspects of Poisson-Lie duality has been taken in \cite{Tyurin:1996bu}.
There a path integral formulation of this duality is presented and more 
importantly the transformation of the dilaton field
is found.  It is worth noticing that Abelian and non-Abelian dualities 
are particular cases of Poisson-Lie T-duality.
The reversibility of Poisson-Lie duality transformations is explicit 
in this construction and resolves, therefore, the problem 
encountered in the case of non-Abelian duality.

It is not sufficient, however, to formulate duality transformations 
between non-linear sigma models at the classical level.
A treatment of these dualities at the quantum level must be carried out.  
This is the only way to find out whether the 
two string theories corresponding to these two sigma models are equivalent.  
If the quantum aspects of Abelian duality are by now 
well understood, the situation regarding  non-Abelian duality has remained, 
until recently \cite{Bossard:2001xq}, unclear. 
Its study has revealed some intriguing issues:  
It was noticed in \cite{Gasperini:1993nz} that the beta functions
of two particular sigma models related by non-Abelian duality are not the same 
and the two models are, thus, quantum mecanically inequivalent.
A global investigation of the origin of this problem has been carried out in \cite{aagl}. 
There, it has been shown that the original
model and its dual are inequivalent, whenever the structure constants of the Lie algebra, 
generated by the Killing vectors, have 
non-vanishing traces. Subsequently, it has been explicitly shown \cite{tyurin}, 
using supersymmetry as a computational tool, that, 
if the original sigma model is conformally invariant, then its dual has vanishing 
Weyl anomaly coefficients too.
More recently, the functional relation (first investigated in \cite{Haagensen} 
for Abelian duality) between the Weyl anomaly 
coefficients of the original model and its dual, under both Abelian and non-Abelian 
dualities, has been formally 
proven \cite{Balog}. This is found by using, at the level of the sigma model, 
a formal path integral procedure to implement  
Abelian and non-Abelian dualities. The desire to give an independant derivation 
of this functional relation by a less formal 
approach has motivated our previous paper \cite{Bossard:2001xq}. Our strategy 
has been inspired firstly, by the reduction of the 
string effective action in the presence of isometries 
\cite{Maharana,kaloper1,kaloper2} 
and secondly by interesting investigations regarding
T-duality beyond the one loop level \cite{Loops,Parsons}. 
Our central tool in showing the equivalence of the two string 
effective actions corresponding to the original sigma model and 
to its dual under non-Abelian duality, has been the use of
Kaluza-Klein decomposition of the different string backgrounds. 
This allows for a much simpler formulation of non-Abelian T-duality
transformations. We have shown that a functional relation holds 
between the Weyl anomaly coefficients of the two models
regardless of the conformal properties of the original theory. 

The aim of this paper is to carry out a quantum analysis of Poisson-Lie T-duality. 
This duality is a canonical transformation and
two sigma models related by Poisson-Lie duality are, therefore, 
equivalent at the classical level \cite{Sfetsos:1998pi,Alvarez:2000bh}. 
In the literature, only few examples of sigma models related by 
Poisson-Lie duality have been treated at the quantum 
level \cite{Alekseev:1996ym,Lledo:1998jm,Sfetsos:1998kr}.
However, no studies exist for a general 
Poisson-Lie dualizable sigma model. This problem will be tackeled here
at the one loop level. We will employ the techniques developed 
in \cite{Bossard:2001xq}, in the case of non-Abelian duality,
to examine the behaviour of the string effective action under 
Poisson-Lie duality. This treatment is valid for both critical and 
non-critical bosonic strings. The use of Kaluza-Klein decompositions 
of the string backgrounds is essential here.  

The outline of this paper is as follows: In section 2 we recall the origin 
of the Poisson-Lie T-duality transformations at the 
level of the non-linear sigma model and introduce our notations. 
In order to cast Poisson-Lie duality transformations in
a symmetrical form, we are naturally led to the introduction of ``intermediate'' 
background fields. A Kaluza-Klein 
reparametrization of the different backgrounds involved in the analysis, 
enables us to find a simple form for the
Poisson-Lie duality transformations. The use of this decomposition in the 
reduction of the low energy effective action
of string theory is the subject of section 3. It is shown that a sufficient 
condition for the  invariance, under Poisson-Lie
duality,  of the reduced string effective action, is the vanishing of the 
traces of the structure constants of each Lie algebra
constituting the Drinfeld double. In section 4 we explain how our results 
concerning the invariance of the string effective action 
invariance, combined with the calculations made in \cite{Bossard:2001xq}, 
can be used to extract the expected relations between the Weyl anomaly 
coefficients of the original model and 
those corresponding to its dual. To illustrate our results,  
we apply our formalism to two explicit examples in section 5. 
Finally, in section 6, we present our conclusions and sketch possible developments of this work.   
Useful identities and the details of the computation are left to an appendix.

\section{ Poisson-Lie Duality }

In this section we recall the main features of the Poisson-Lie T-duality
at the level of the sigma model and
introduce new redefinitions to make the duality transformations
more symmetrical. 
Poisson-Lie duality is based on a Drinfeld double ${\cal D}$
corresponding to two Lie algebras ${\cal G}$ and $\t{\cal G}$.
The generators of ${\cal G}$ and $\t{\cal G}$ are denoted respectively
$T_a$ and $\t{T}^a$ and satisfy the commutation relations
\bea
[T_a , T_b] = f_{ab}^c T_c,\hspace{20mm} [\tilde{T}^a ,\tilde{T}^b] 
=\tilde{f}^{ab}_c \tilde{T}^c, \nonumber
\eea
\bea
[T_a , \tilde{T}^b] =\tilde{f}^{bc}_a T_c -  f_{ac}^b \tilde{T}^c.
\eea
The structure constants $f^a_{bc}$ and $\t{f}_a^{bc}$ are subject to
the Jacobi identities
\be
f^e_{ab} \t{f}_e^{cd} = 
f^c_{ea} \t{f}_b^{de} - f^d_{ea} \t{f}_b^{ce}
-f^c_{eb} \t{f}_a^{de} + f^d_{eb} \t{f}_a^{ce}\,\,.
\label{ff}
\ee 
The Drinfeld double ${\cal D}$ is equipped 
with an invariant inner product $<,>$ with the 
following properties
\bea
<T_a,T_b>=<\t{T}^a,\t{T}^b>=0\,\,,\;\;\;<T_a,\t{T}^b>=\delta_a^b\,\,\,,
\label{bilinear_product1}
\eea
and
\bea
<l T_A l^{-1}, T_B>=< T_A, l^{-1} T_B l>\,\,,
\label{bilinear_product2}
\eea
where $T_A$ stands for $T_a$ or $\t{T}^a$ and $l$ is an element of the Lie group 
$D$ corresponding to ${\cal D}$.
The following definitions are also needed
\bea
g^{-1} T_a g &=& {a(g)}_a^{\;\;\;b}T_b\,\,,\nonumber \\
g^{-1} \t{T}^a g &=& {b(g)}^{ab} T_b + {a^{-1}(g)}_b^{\;\;\;a} \t{T}^b\,\,, 
\nonumber \\
{\Pi(g)}^{ab} &=& {b(g)}^{ca} {a(g)}_c^{\;\;\;b}\,\,,
\label{def_pi}
\eea 
where $g$ is an element of the Lie group corresponding to 
the Lie algebra $\cal G$.

The original sigma model is defined by the action 
\bea
S=\int d\sigma d\bar{\sigma}
\Big[ P_{\mu\nu}\partial x^{\mu} \bar{\partial} x^{\nu}
+P^{(1)}_{\;\;\;a \nu} (g^{-1} \partial g)^a \bar{\partial} x^{\nu}
+P^{(2)}_{\;\;\;\mu b} \partial x^{\mu} (g^{-1} \bar{\partial} g)^b \nonumber \\
+E_{ab} (g^{-1} \partial g)^a  (g^{-1} \bar{\partial} g)^b 
-\frac{1}{4}R^{(2)}\Ph \Big]\,\,,
\label{smodelaction}
\eea
where $R^{(2)}$ is the curvature of the worldsheet.
Since the background fields $(P,P^{(1)},P^{(2)},E,\Ph)$
in this action depend a priori on the group elements $g$,
the model, in general, has no isometries. 
This lack of symmetry is typical of Poisson-Lie duality.
The backgrounds appearing in this action are given in matrix notation
by \cite{Tyurin:1996bu,Jafarizadeh:1999xv}
\bea
P=\tilde{F}-\tilde{F}^{(2)} 
\Pi E \tilde{Q}^{-1} \tilde{F}^{(1)}, \;\;\;
P^{(1)}=E \tilde{Q}^{-1} \tilde{F}^{(1)}, \;\;\; 
P^{(2)}=\tilde{F}^{(2)} \tilde{Q}^{-1} E, \;\;\; \nonumber \\
E=(\tilde{Q}^{-1}+\Pi)^{-1}, \;\;\; 
\Ph= \Pht_0 -\ln \det \tilde{Q}^{-1} + \ln \det E\,\,.
\label{PEoriginal}
\eea
The matrices 
$(\tilde{F}, \tilde{F}^{(1)}, \tilde{F}^{(2)}, 
\tilde{Q})$ and the scalar field
$\Pht_0$ are all functions of the 
variables $x^{\mu}$ only.

The dual sigma model is given by
\bea
\tilde{S}=\int d\sigma d\bar{\sigma}
\Big[\tilde{P}_{\mu\nu}\partial x^{\mu} \bar{\partial} x^{\nu}
+\tilde{P}^{(1)a}_{\;\;\;\;\;\;\nu} (\gt^{-1} \partial \gt)_a 
\bar{\partial} x^{\nu}
+\tilde{P}_{\;\;\;\mu}^{(2)\;\;\;b} \partial x^{\mu} 
(\gt^{-1} \bar{\partial} \gt)_b \nonumber \\
+\tilde{E}^{ab} (\gt^{-1} \partial \gt)_a  (\gt^{-1} \bar{\partial} \gt)_b 
-\frac{1}{4}R^{(2)}\Pht\Big]\,\,,
\eea
where $\t{g}$ is an element of the Lie group corresponding
to $\t{\cal G}$. 
The backgrounds of the dual theory are related to those of the
original one by
\bea
\tilde{P}=\tilde{F}-\tilde{F}^{(2)} \tilde{E} \tilde{F}^{(1)}, \;\;\;
\tilde{P}^{(1)}=\tilde{E} \tilde{F}^{(1)}, \;\;\; 
\tilde{P}^{(2)}=-\tilde{F}^{(2)} \tilde{E},\;\;\; \nonumber \\
\tilde{E}=(\tilde{Q}+\tilde{\Pi})^{-1}, \;\;\;
\Pht=\Pht_0 + \ln \det \tilde{E}\,\,,
\label{PEdual}
\eea
where $\t{\Pi}(\t{g})_{ab}=
{\t{b}(\t{g})}_{ca} {\t{a}(\t{g})}^c_{\;\;\;b}$ 
is defined as in (\ref{def_pi}) by replacing
untilded quantities by tilded ones and vice versa.

The transformations for the two dilatons $\Ph$ and $\t{\Ph}$
have been obtained in \cite{Tyurin:1996bu} by quantum considerations. 
These calculations were based on a regularization of a functional determinant  
in a path integral formulation of Poisson-Lie duality.
This is consistent with previous transformations of the dilaton
obtained for Abelian and non-Abelian T-dualities \cite{dilaton}.

Notice, however, that the Poisson-Lie duality transformations
(\ref{PEoriginal}) and (\ref{PEdual})
are not explicitly symmetric.
Namely, one would like to pass from one group of relations
to the other simply by replacing untilded symbols by tilded
ones and vice versa.
A remedy to this can be found by introducing the following quantities
\bea 
Q=\t{Q}^{-1}\,\, , \,\, 
F^{(1)}=Q\t{F}^{(1)}\,\, , \,\, 
F^{(2)}=-\t{F}^{(2)}Q\,\, ,  \nonumber \\
F=\t{F}-\t{F}^{(2)}Q\t{F}^{(1)}\,\, , \,\,
\Ph_0=\Pht_0+\ln \det Q \,\,.
\label{QFrelations}
\eea
In terms of these new tensors, the backgrounds of the original theory
are given by
\bea
P=F-F^{(2)} E F^{(1)}, \;\;\;
P^{(1)}=E F^{(1)}, \;\;\; 
P^{(2)}=-F^{(2)} E,\;\;\; \nonumber \\
E=(Q+\Pi)^{-1}, \;\;\;
\Ph=\Ph_0 + \ln \det E\,\,.
\label{PEoriginal2}
\eea
One sees, indeed, that the backgrounds of the dual (\ref{PEdual})
are now obtained from (\ref{PEoriginal2})
by simply replacing untilded quantities by tilded ones.
This becomes a crucial point when dealing with the string effective actions
corresponding to the two sigma models.
Notice that if one takes a dual Abelian group ($\t{f}_c^{ab}=0$) 
then one finds 
\bea
\Pi^{ab}=0\,\,,\t{\Pi}_{ab}=y_c f^c_{ab}\,\,,
\eea
where $y_a$ are local coordinates characterizing the group element $g$.
We have in this case
\bea
\t{E}^{ab}=(E_{ab}+y_c f^c_{ab})^{-1}\,\,,
\eea
recovering thus the usual non-Abelian duality.
Similarly, choosing the original group to be Abelian
leads then to a symmetric version of non-Abelian duality.
It is amusing to notice that the redefinitions (\ref{QFrelations})
have exactly the same form as the Abelian T-duality transformations
relating a sigma model with backgrounds $(F,F^{(1)},F^{(2)},Q,\Ph_0)$
to its dual with backgrounds $(\t{F},\t{F}^{(1)},\t{F}^{(2)},\t{Q},\Pht_0)$.
Furthermore, the expressions in (\ref{PEoriginal2}) are of
the same form as non-Abelian T-duality transformations
except that the $\Pi^{ab}$ term replaces the $y^c \t{f}_c^{ab}$
of non-Abelian duality.

Since the expressions of the backgrounds of the original and dual
theories are now explicitly symmetric under the exchange of tilded
and untilded quantities, we will deal only with the original
theory in what follows.
In order to address the question of the invariance of the string
effective action under Poisson-Lie duality transformations,  
we need to rewrite the action (\ref{smodelaction}) in the usual standard form
\bea
S=\int d\sigma d\bar{\sigma} \Big[
\big(G_{MN}+B_{MN}\big)\partial X^M \bar{\partial} X^N
-\frac{1}{4}R^{(2)}\Ph\Big]\,\,,
\eea
where $X^M=(x^{\mu},y^i)$ and 
$y^i$ are local coordinates on the group manifold
corresponding to $\cal{G}$.
The metric $G_{MN}$, the antisymmetric tensor $B_{MN}$ and the dilaton $\Ph$
are then easily identified.
The vielbeins are introduced through 
$(g^{-1}\partial g)^a=e^a_i(y)\partial y^i$ 
and verify
\be
\partial _{i}e_{j}^{a}-\partial _{j}e_{i}^{a}=-f_{bc}^{a}e_{i}^{b}e_{j}^{c}\,\,.
\label{fee}
\ee
In analogy with the non-Abelian duality case \cite{Bossard:2001xq}
we use a Kaluza-Klein decomposition of the different backgrounds.
In this decomposition, the Poisson-Lie duality transformations
take a simpler form.
It is convenient for this purpose to introduce the following notations
\bea
E_{ab}=S_{ab}+A_{ab}\,\,,\,\,Q^{ab}=(S_0)^{ab}+(v_0)^{ab}\,\,,
(A_0)^{ab}=(v_0)^{ab}+\Pi^{ab}\,\,,
\eea 
where $S_{ab}$ and $(S_0)^{ab}$ are symmetric while 
$A_{ab}$, $(v_0)^{ab}$, $(A_0)^{ab}$ and $\Pi^{ab}$ are antisymmetric.
We start by introducing the ``intermediate'' backgrounds as follows:
A symmetric part given by
\bea
(G_0)_{MN} &\equiv&
\left( \begin{array}{cc}
F_{(\mu \nu)} 
& \frac{1}{2}({F^{(2)}}_{\mu}^a+{F^{(1)}}_{\mu}^a)\\
\frac{1}{2}({F^{(2)}}_{\mu}^a+{F^{(1)}}_{\mu}^a)
& Q^{(ab)}
\end{array}\right)
\nonumber \\
&=&
\left( \begin{array}{cc}
g_{\mu \nu }+(S_0)^{ab}t_{\mu a}t_{\nu b} 
& t_{\mu a}(S_0)^{ab}\\
t_{\mu a}(S_0)^{ab} & (S_0)^{ab}
\end{array}\right)\,\,,
\label{G_0}
\eea
and an antisymmetric part 
\bea
(B_0)_{MN} &\equiv&
\left( \begin{array}{cc}
F_{[\mu \nu]} 
& \frac{1}{2}({F^{(2)}}_{\mu}^a-{F^{(1)}}_{\mu}^a)\\
-\frac{1}{2}({F^{(2)}}_{\mu}^a-{F^{(1)}}_{\mu}^a)
& Q^{[ab]}
\end{array}\right)
\nonumber \\
&=& 
\left( \begin{array}{cc}
b_{\mu\nu}
-\frac{1}{2}(t_{\mu a}u_{\nu}^a-t_{\nu a}u_{\mu}^a)
\,\,\,& u_{\mu}^a \\
-u_{\mu}^a & (v_0)^{ab}
\end{array}\right) \,\,.
\label{B_0}
\eea  
Notice that all the components here are functions of 
the coordinates $x^{\mu}$ only.

The above decomposition allows one to reparametrize
the metric $G_{MN}$ as
\begin{equation}
G_{MN}=\left( \begin{array}{cc}
g_{\mu \nu }+h_{ij}V_{\mu }^{i}V_{\nu }^{j} \,\,\,& V^{i}_{\mu }h_{ij}\\
V^{i}_{\mu }h_{ij} & h_{ij}
\end{array}\right)\,\, . 
\label{G_MN}
\end{equation}
while the antisymmetric tensor $B_{MN}$ can be decomposed as
\begin{equation}
B_{MN}=\left( \begin{array}{cc}
b_{\mu \nu }-\frac{1}{2}(V^{k}_{\mu }B_{\nu k}-V^{k}_{\nu }B_{\mu k}) 
\,\,\,& B_{\mu i}\\
-B_{\mu i} & b_{ij}
\end{array}\right)\,\,,
\label{B_MN} 
\end{equation}
where 
\bea
V_{\mu}^k&=&\Big[t_{\mu a} (A_0)^{ab} -u_{\mu}^b\Big](e^{-1})_b^k\,\,,
\nonumber \\
B_{\mu k}&=&-\Big[t_{\mu a} (S_0)^{ab} S_{bc} +u_{\mu}^a A_{ac}\Big]e^c_k\,\,,
\nonumber \\
h_{ij}&=&e_i^a S_{ab} e_j^b\,\,,
\nonumber \\
b_{ij}&=&e_i^a A_{ab} e_j^b\,\,,
\nonumber \\
E_{ab}&=&S_{ab}+A_{ab}=\Big((S_0)^{ab}+(A_0)^{ab}\Big)^{-1}\,\,.
\label{omegalambda}
\eea
From the last equation in (\ref{omegalambda}), important relations are derived
\bea
S_{ab}(S_0)^{bc}+A_{ab}(A_0)^{bc}=\delta_a^c \;\;,\;
S_{ab}(A_0)^{bc}+A_{ab}(S_0)^{bc}=0\,\,.
\label{fundamental1}
\eea
They will be fundamental in the reduction of the effective action
in the next section.

Similarly, by decomposing 
$(\t{F},\t{F}^{(1)},\t{F}^{(2)},\t{Q})$ 
as in (\ref{G_0},\ref{B_0}),
the backgrounds of the dual theory have the
Kaluza-Klein reparametrization
\begin{equation}
\t{G}_{MN}=\left( \begin{array}{cc}
\t{g}_{\mu \nu }+\t{h}^{ij}\t{V}_{\mu i}{\t V}_{\nu j} \,\,\,& 
{\t V}_{\mu i}{\t h}^{ij}\\
{\t V}_{\mu i}{\t h}^{ij} & {\t h}^{ij}
\end{array}\right)\,\,, 
\label{tG_MN}
\end{equation}
and
\begin{equation}
\t{B}_{MN}=\left( \begin{array}{cc}
\t{b}_{\mu \nu }
-\frac{1}{2}(\t{V}_{\mu k}\t{B}_{\nu}^k-\t{V}_{\nu k}\t{B}_{\mu}^k) 
\,\,\,& \t{B}_{\mu}^i\\
-\t{B}_{\mu}^i & \t{b}^{ij}
\end{array}\right)\,\,,
\label{Bt_MN} 
\end{equation}
where 
\bea
\t{V}_{\mu k}
&=&
\Big[\t{t}_{\mu}^a {(\t{A}_0)}_{ab} -\t{u}_{\mu b}\Big](\t{e}^{-1})^b_k\,\,,
\nonumber \\
\t{B}_{\mu}^k
&=&
-\Big[\t{t}_{\mu}^a {(\t{S}_0)}_{ab} \t{S}^{bc} 
+\t{u}_{\mu a} \t{A}^{ac}\Big]\t{e}_c^k\,\,,
\nonumber \\
\t{h}^{ij}&=&\t{e}_a^i \t{S}^{ab} \t{e}_b^j\,\,,
\nonumber \\
\t{b}^{ij}&=&\t{e}_a^i \t{A}^{ab} \t{e}_b^j\,\,,
\nonumber \\
\t{E}^{ab}&=&\t{S}^{ab}+\t{A}^{ab}
=\Big({(\t{S}_0)}_{ab}+{(\t{A}_0)}_{ab}\Big)^{-1}\,\,.
\label{omegatlambdat}
\eea
Notice that the dual relations of (\ref{fundamental1}) can be derived 
from the last equation of (\ref{omegatlambdat}).

In this decomposition, the Poisson-Lie duality transformations
take the following simple form
\bea
g_{\mu \nu}&=&\t{g}_{\mu \nu}\,\,,\nonumber \\
b_{\mu \nu}&=&\t{b}_{\mu \nu}\,\,,\nonumber \\
{u'}^a_{\mu} &=& -{\t{t}}^a_{\mu}\,\,, \nonumber \\
t_{\mu a} &=& -{\t{u}'}_{\mu a}\,\,, \nonumber \\ 
{(S_0)}^{ab}+{(v_0)}^{ab} &=& \Big({(\t{S}_0)}_{ab}+{(\t{v}_0)}_{ab}\Big)^{-1}\,\,,
\label{lambdalambdatilde}
\eea
where we have defined 
\bea
{u'}^a_{\mu}&=&u^a_{\mu}-t_{\mu b}{(v_0)}^{ba}\,\,, 
\nonumber \\
{\t{u}'}_{\mu a}&=&\t{u}_{\mu a}-\t{t}_{\mu}^b{(\t{v}_0)}_{ba}\,\,.
\eea
The last equation of the Poisson-Lie duality relations (\ref{lambdalambdatilde}) implies
\bea
{(S_0)}^{ab}{(\t{S}_0)}_{bc}+{(v_0)}^{ab}{(\t{v}_0)}_{bc}=\delta^a_c \;\;,\;
{(S_0)}^{ab}{(\t{v}_0)}_{bc}+{(v_0)}^{ab}{(\t{S}_0)}_{bc}=0\,.
\label{fundamental2}
\eea
Again, these relations will be useful in showing
the invariance of the reduced string effective action under Poisson-Lie duality.

As is well known, 
the Kaluza-Klein reparametrization has other important advantages. 
Firstly, it enables one to calculate the inverse metric 
straightforwardly and, consequently, the scalar curvature.
In our case, the inverse of $G_{MN}$ is 
\be
G^{MN}=\left( \begin{array}{ll}
g^{\mu \nu } \,\,\,& -V^{\mu i} \\
-V^{\mu i} & h^{ij}+V^{i}_{\mu }V^{\mu j}
\end{array}\right) 
\label{Ginv}
\ee
where Greek indices are raised and lowered with the metric $g_{\mu \nu}$
while latin indices are raised and lowered with $h_{ij}$ 
whose inverse is $h^{ij}$. 
Secondly, the determinant of the metric $G_{MN}$, 
is simply given by
\begin{equation}
\det G_{MN}=\det g_{\mu \nu }\det h_{ij}\,\,.
\label{determinant}
\end{equation}
These last two properties will be useful in the next sections.

In order to complete our decomposition of the backgrounds,
we deal now with the reparametrization of the dilaton field.
Recall that the dilaton in the original theory is given by
$\Ph=\Ph_0+\ln \det E$ and its counterpart in the dual theory is written
as $\Pht=\Pht_0 +\ln \det \t{E}$ 
with $\Ph_0$ and $\Pht_0$ related by
$\Ph_0=\Pht_0+\ln \det Q$.
Now, using the fundamental relations (\ref{fundamental1})
one can shown that $E S_0= S {(E^{-1})}^t$ where $t$ denotes the 
transpose operation. 
As a consequence we have
\bea
\ln \det E=\frac{1}{2} \ln \det S -\frac{1}{2} \ln \det S_0
\label{lndetE}
\eea
It is convenient to use the following decompositions for the dilaton field
\be
\Ph(x,y)=\psi(x,y)+\theta(x,y) , \,\, \Ph_0(x)=\psi_0(x)+\theta_0(x)\,\,,
\label{decompodilaton1}
\ee
where $\theta=\frac{1}{2} \ln \det h_{ij}$ and $\theta_0=\frac{1}{2} \ln \det S_0$.
The dilaton relation from (\ref{PEoriginal2}) then gives 
\be 
\psi=\psi_0(x)-\ln \det e(y)\,\,.
\label{psi_prop}
\ee
Similarly, the dilaton in the  dual theory is decomposed as
\be
\Pht(x,y)=\t{\psi}(x,y)+\t{\theta}(x,y)\,\,,\,\, 
\Pht_0(x)=\t{\psi}_0(x)+\t{\theta}_0(x)\,\,,
\label{decompodilaton2}
\ee
where $\t{\theta}=\frac{1}{2} \ln \det \t{h}^{ij}$ 
and $\t{\theta}_0=\frac{1}{2} \ln \det \t{S}_0$.
In the same way, the dilaton expresson in (\ref{PEdual}) leads to
\be 
\t{\psi}=\t{\psi}_0(x)-\ln \det \t{e}(y)\,\,.
\label{psit_prop}
\ee
Furthermore, from the fundamental relations (\ref{fundamental2}), 
one shows that $Q \t{S}_0= S_0 \t{Q}^t$
which in turn yields
\bea
\ln \det Q=\frac{1}{2} \ln \det S_0 -\frac{1}{2} \ln \det \t{S}_0\,\,.
\eea
Consequently, the expression $\Ph_0=\Pht_0+\ln \det Q$, gives
\be
\psi_0=\t{\psi}_0\,\,.
\ee
All these relations will be our main tools in the next section.

\section{ String Effective Action }

In this section, we deal with our main concern in this paper, 
namely the invariance under Poisson-Lie
T-duality of the effective action of bosonic string theory. 
It is well known that the string effective action 
is connected to the two dimensional non-linear sigma model
through the Weyl anomaly coefficients
\bea
\bar{\beta }^{G}_{MN} & = & R_{MN}+\nabla _{M}\partial _{N}\varphi 
-\frac{1}{4}H_{MPQ}{H_N}^{PQ} \,\, , \nonumber \\
\bar{\beta }^{B}_{MN} & = & -\frac{1}{2}\nabla ^{P}H_{MNP}
+\frac{1}{2}H_{MNP}\partial ^{P}\varphi \,\, ,\nonumber \\
\bar{\beta }^{\varphi } & = & -\frac{1}{4}\nabla ^{2}\varphi 
+\frac{1}{4}\partial _{P}\varphi \partial ^{P}\varphi -\frac{1}{24}H_{MNP}H^{MNP}
-\frac{\Lambda}{4}\,\,,
\eea
where $\Lambda$ is a cosmological constant which vanishes for critical strings.
Our analysis applies also to non-critical strings, 
i.e. when $\Lambda$ is different from zero. 
Hence, we will keep $\Lambda$ throughout this paper.
The Weyl anomaly coefficients (the beta functions) 
can be derived as the equations of motion of the string effective action
\bea
\Gamma[G,B,\Ph]=\int d^d x d^n y \sqrt{G}e^{-\Ph} L \,\, ,
\label{effectiveaction}
\eea
where $L$ is given by
\bea
L=R+2\nabla_M \partial^M \Ph +\partial_{M}\Ph \partial ^{N}\Ph 
-\frac{1}{12}H_{MNP}H^{MNP}+\Lambda\ \,\, .
\eea
In this expression, $R$ is the scalar curvature of the metric $G_{MN}$ and
$H_{MNP}$, defined by 
$H_{MNP}=\partial _{M}B_{NP}+\partial _{N}B_{PM}+\partial _{P}B_{MN}$,
is the torsion of the antisymmetric field $B_{MN}$ 
while $\Ph$ is the dilaton field.
We have chosen, at this stage, to make an integration by parts
in the dilatonic part of $L$. This will spare us later integrations by parts 
when showing the invariance of $L$ under Poisson-Lie T-duality transformations.
It is understood that a similar Lagrangian $\t{L}[\t{G},\t{B},\Pht]$
is defined for the backgrounds of the dual sigma model.

Let us first explain our strategy in reaching our goal:
We start by reducing, with the help of the Kaluza-Klein decompositions
of the previous section, the string effective action $\Gamma[G,B,\Ph]$
corresponding to the original sigma model.
On the other hand, 
the reduction of the string effective action $\t{\Gamma}[\t{G},\t{B},\Pht]$ corresponding to
the dual sigma model is obtained from that of $\Gamma[G,B,\Ph]$ by simply
replacing untilded quantities by tilded ones and vice versa.
This is due to our symmetric formulation of Poisson-Lie duality.
Finally, the expressions of the reduced actions $\Gamma$ and 
$\t{\Gamma}$ are compared using the Poisson-Lie duality relations (\ref{lambdalambdatilde}).

The first result concerns the weight factor $\sqrt{G}e^{-\Ph}$
in the action (\ref{effectiveaction}).
The different dilatonic properties of the previous 
section and the determinant property (\ref{determinant})
lead to 
\bea
\sqrt{G}e^{-\Ph}=\det e \sqrt{g}e^{-\psi_0}\,\,.
\eea
In this expression the dependence on $y^i$ is only in $\det e$. 
Similarly, the weight factor in the string effective action
corresponding to the dual theory, is given by
\bea
\sqrt{\t{G}}e^{-\Pht}=\det \t{e} \sqrt{g}e^{-\psi_0}\,\,.
\eea
Therefore, the two integration measures in the two theories are equal up to
the determinants of the vielbeins $\det e$ and $\det \t{e}$.
We will show, in what follows, that $L$ and $\t{L}$ are both
independant of the coordinates $y^i$. 
As a consequence, the two effective actions $\Gamma$ and $\t{\Gamma}$
are equal up to the volume elements $\int d^n y \det e$ and  $\int d^n y \det \t{e}$.
 
With the help of the Kaluza-Klein decomposition of the metric and of the
antisymmetric tensor given in the previous section we write explicitly the different
contributions to $L$.
First, the Kaluza-Klein decomposition of the metric $G_{MN}$ 
yields the Ricci scalar \cite{Bossard:2001xq}
\begin{eqnarray}
R(G) & = & R(g)
\nonumber \\
 && +\{
h^{ik}h^{jl}\partial _{i}\partial _{j}h_{kl}
-h^{ij}h^{kl}\partial _{i}\partial _{j}h_{kl}
 \nonumber \\
 && -\frac{1}{2}h^{ir}h^{jl}h^{ks}\partial _{i}h_{jk}\partial _{l}h_{rs}
+\frac{3}{4}h^{il}h^{jr}h^{ks}\partial _{i}h_{jk}\partial _{l}h_{rs} 
\nonumber \\
 && -h^{ij}h^{kr}h^{ls}\partial _{i}h_{jk}\partial _{l}h_{rs}
-\frac{1}{4}h^{il}h^{jk}h^{rs}\partial _{i}h_{jk}\partial _{l}h_{rs}
+h^{ij}h^{kl}h^{rs}\partial _{i}h_{jk}\partial _{l}h_{rs}
\} 
\nonumber\\
 && +g^{\mu \nu }\{
(-h^{ij}\nabla _{\mu }\nabla _{\nu }h_{ij}
+\frac{3}{4}h^{ik}h^{jl}\nabla _{\mu }h_{ij}\nabla _{\nu }h_{kl}
-\frac{1}{4}h^{ij}h^{kl}\nabla _{\mu }h_{ij}\nabla _{\nu }h_{kl})
\nonumber  \\
 && +(2\partial _{i}\nabla _{\mu }V_{\nu }^{i}
+h^{ik}\partial _{i}V_{\nu }^{j}\nabla _{\mu }h_{jk}
+2h^{jk}V^{i}_{\mu }\nabla _{\nu }\partial _{i}h_{jk}
+h^{jk}\partial _{i}h_{jk}\nabla _{\mu }V^{i}_{\nu } \nonumber \\
 && +h^{jk}\partial _{i}V_{\mu }^{i}\nabla _{\nu }h_{jk}
-\frac{3}{2}h^{jl}h^{kr}V^{i}_{\mu }\nabla _{\nu }h_{lr}\partial _{i}h_{jk}
+\frac{1}{2}h^{jk}h^{lr}V^{i}_{\mu }\nabla _{\nu }h_{lr}\partial _{i}h_{jk}) \nonumber \\
 && +(-2V_{\mu }^{i}\partial _{i}\partial _{j}V_{\nu }^{j}
-\frac{1}{2}\partial _{i}V^{j}_{\mu }\partial _{j}V^{i}_{\nu }
-\partial _{i}V^{i}_{\mu }\partial _{j}V^{j}_{\nu }
-\frac{1}{2}h^{ik}h_{jl}\partial _{i}V^{j}_{\mu }\partial _{k}V_{\nu }^{l} \nonumber \\
 && -h^{il}V^{k}_{\mu }\partial _{i}V^{j}_{\nu }\partial _{k}h_{jl}
-h^{kl}V^{i}_{\mu }\partial _{i}V^{j}_{\nu }\partial _{j}h_{kl}
-h^{kl}V^{j}_{\mu }\partial _{i}V^{i}_{\nu }\partial _{j}h_{kl} \nonumber \\
 && -h^{kl}V^{i}_{\mu }V^{j}_{\nu }\partial _{i}\partial _{j}h_{kl}
 +\frac{3}{4}h^{jr}h^{ks}V^{i}_{\mu }V^{l}_{\nu }\partial _{i}h_{jk}\partial _{l}h_{rs} 
-\frac{1}{4}h^{jk}h^{rs}V^{i}_{\mu }V^{l}_{\nu }\partial _{i}h_{jk}
\partial _{l}h_{rs})\} \nonumber \\
 && -\frac{1}{4}g^{\mu \rho}g^{\nu \lambda}h_{ij}
(V^{i}_{\mu \nu}-{}F^{i}_{\mu \nu})
(V^{j}_{\rho \lambda}-{}F^{j}_{\rho \lambda})
\end{eqnarray}
with the definitions 
\begin{equation}
V^{i}_{\mu \nu} =  
\partial_{\mu } V^{i}_{\nu}
-\partial_{\nu} V^{i}_{\mu}
\,\,\,,\;\;\;
F^{i}_{\mu \nu}  =  
V_{\mu}^{k} \partial_k V_{\nu}^{i}
-V_{\nu}^{k}\partial_k V_{\mu}^{i}\,\,.
\end{equation}
The terms have been assembled according to the number of factors of 
$g^{\mu\nu}$ and the number of factors of $V_\mu^i$. 
This separation of terms will serve as a guide in our calculation.
Since $g_{\mu\nu}$ is invariant under Poisson-Lie T-duality 
($g_{\mu \nu}=\t{g}_{\mu \nu}$) then 
$R(g)$ is obviously invariant too.

Second, we develop the $H_{MNP}H^{MNP}$ term.
Using the decomposition of $G_{MN}$ in (\ref{Ginv}),
it can be shown that
\begin{equation}
H_{MNP}H^{MNP}=(h_{\mu \nu \rho }h^{\mu \nu \rho })+3(h_{\mu \nu i}h^{\mu \nu i})
+3(h_{\mu ij}h^{\mu ij})+(h_{ijk}h^{ijk})
\label{hh}
\end{equation}
where we raise Greek indices with $g^{\mu \nu }$ and Latin
indices with $h^{ij}={(e^{-1})}_{a}^{i}S^{ab}{(e^{-1})}^{j}_{b}$ 
with $S^{ab}$ the inverse of $S_{ab}$.
Here the components of $h_{MNP}$ are defined by
\begin{eqnarray}
h_{ijk} & = & H_{ijk}\nonumber \\
h_{\mu ij} & = & H_{\mu ij}-V^{k}_{\mu }H_{kij}\nonumber \\
h_{\mu \nu i} & = & H_{\mu \nu i}-\{V^{k}_{\mu }H_{k\nu i}
-(\mu \leftrightarrow \nu )\}+V^{k}_{\mu }V^{l}_{\nu }H_{kli}\nonumber \\
h_{\mu \nu \rho } & = & H_{\mu \nu \rho }-\{V^{k}_{\mu }H_{k\nu \rho }
+\mathbf{c.p.}\}+\{V^{k}_{\mu }V^{l}_{\nu }H_{kl\rho }
+\mathbf{c.p.}\}-V^{k}_{\mu }V^{l}_{\nu }V^{m}_{\rho }H_{klm}
\label{h_MNP} 
\end{eqnarray}
where $\mathbf{c.p.}$ stands for cyclic permutations.
Now, putting the decomposition (\ref{B_MN}) of the antisymmetric tensor $B_{MN}$
into these expressions, one finds 
\begin{eqnarray}
h_{ijk} & = & \partial _{i}b_{jk}+\mathbf{c}.\mathbf{p}.\\
h_{\mu ij} & = & (\partial _{\mu }b_{ij})+(-V^{k}_{\mu }\partial _{k}b_{ij}
+\{\partial _{j}B_{\mu i}'+(\partial _{j}V^{k}_{\mu })b_{ki}-(i\leftrightarrow j)\})\\
h_{\mu \nu i} & = & (B_{\mu \nu i}'+V_{\mu \nu }^{k}b_{ki})
+([V^{k}_{\mu }B_{i\nu k}'-(\mu \leftrightarrow \nu )]
-\frac{1}{2}\partial _{i}U_{\mu \nu }-{}F^{k}_{\mu \nu }b_{ki})\\
h_{\mu \nu \rho } & = & (\partial _{\rho }b_{\mu \nu })
+(-V_{\rho }^{k}\partial _{k}b_{\mu \nu })
+\frac{1}{2}(V^{k}_{\mu \rho }B_{\nu k}'+B_{\mu \rho k}'V_{\nu }^{k})\nonumber \\
 &  & 
+\frac{1}{2}(F^{k}_{\rho \mu }B_{\nu k}'+V_{\rho }^{k}V_{\mu }^{l}B_{l\nu k}')+{\bf c.p.}
\end{eqnarray}
where we have defined
\begin{eqnarray}
B_{\mu i}' & = & B_{\mu i}-V^{k}_{\mu }b_{ki}\nonumber\\
B_{\mu \rho i}' & = & \partial _{\mu }B_{\rho i}'-\partial _{\rho }B_{\mu i}'
\nonumber\\
B_{i\rho j}' & = & \partial _{i}B_{\rho j}'-\partial _{j}B_{\rho i}'
\nonumber\\
U_{\mu \nu } & = & V^{k}_{\mu }B_{\nu k}'-V^{k}_{\nu }B_{\mu k}'\,\,.
\end{eqnarray}
The introduction of these new tensors is motivated by 
their simple expressions,  
as given in the appendix A, in terms of ($t,u,S_0,v_0,...$).
In terms of these new tensors, we find
\begin{eqnarray}
-\frac{1}{12}H_{MNP}H^{MNP}&=&	
\{\frac{1}{2}h^{km}h^{il}h^{jn}\partial _{k}b_{ij}\partial _{l}b_{mn}
-\frac{1}{4}h^{kl}h^{im}h^{jn}\partial _{k}b_{ij}\partial _{l}b_{mn}\}
\nonumber\\
&&+ g^{\mu\nu}h^{im}h^{jn}
\{-\frac{1}{4}\partial_{\mu}b_{ij}\partial_{\nu}b_{mn}
+(\partial_{\mu}b_{ij}\partial_{m}B_{\nu n}'
+\partial_{\mu}b_{ij}\partial_{m}V^{k}_{\nu}b_{kn}
\nonumber\\
&&+\frac{1}{2}
\partial_{\mu}b_{ij}V^{k}_{\nu}\partial_{k}b_{mn})
+(-\frac{1}{4}V^{k}_{\mu }V^{l}_{\nu }
\partial_{k}b_{ij}\partial_{l}b_{mn}
-V^{l}_{\mu }
\partial _{l}b_{ij}\partial _{m}V^{k}_{\nu }b_{kn}
\nonumber\\
&&-V^{l}_{\mu }
\partial _{l}b_{mn}\partial _{i}B_{\nu j}'
-\frac{1}{2}
\partial _{m}V^{l}_{\mu }b_{ln}
[\partial_i V^{k}_{\nu }b_{kj}-(i\leftrightarrow j)]
\nonumber \\
&&-\partial _{m}V^{k}_{\mu}b_{kn}
[\partial _{i}B_{\nu j}'-(i\leftrightarrow j)]
-\frac{1}{2}
\partial _{m}B_{\mu n}'
[\partial _{i}B_{\nu j}'-(i\leftrightarrow j)])\}
\nonumber\\
&&-\frac{1}{4}g^{\mu \rho}g^{\nu \lambda}h^{ij}
(h^{(1)}_{\mu \nu i}+h^{(2)}_{\mu \nu i})
(h^{(1)}_{\rho \lambda j}+h^{(2)}_{\rho \lambda j})
\nonumber\\
&& -\frac{1}{12}g^{\mu\alpha}g^{\nu\beta}g^{\rho\gamma}
h_{\mu\nu\rho}h_{\alpha\beta\gamma}
\label{hcarre}
\end{eqnarray}
where we have defined 
\begin{equation}
h^{(1)}_{\mu\nu i}=B_{\mu\nu i}'+V_{\mu\nu}^{k}b_{ki}
\,\,\,,\,\,\,
h^{(2)}_{\mu\nu i}=[V^{k}_{\mu }B_{i\nu k}'-(\mu \leftrightarrow \nu )]
-\frac{1}{2}\partial _{i}U_{\mu \nu }-F^{k}_{\mu \nu }b_{ki}\,\,.
\end{equation}
Again, the terms of (\ref{hcarre}) have been assembled according 
to the number of factors of $g_{\mu\nu}$ and ``gauge fields'' $V_\mu^i$ and $B'_{\mu i}$.

Finally, we turn our attention to the dilatonic part of $L$.
Using the decompositions (\ref{Ginv}) and (\ref{decompodilaton1})
, the dilatonic contribution to the original effective action is 
\bea
2\nabla_M \partial^M \Ph - \partial_M \Ph \partial^M \Ph 
&=& 
G^{MN}[2\partial_M \partial_N \Ph -2 \Gamma_{MN}^P \partial_P \Ph 
-\partial_M \Ph \partial_N \Ph] \nonumber 
\\
&=& 
(2G^{\mu \nu }\partial _{\mu }\partial _{\nu }\psi 
-G^{\mu \nu }\partial _{\mu }\psi \partial _{\nu }\psi\nonumber\\&& 
+4G^{\mu i}\partial _{i}\partial _{\mu }\psi 
-2G^{MN}\Gamma ^{\lambda }_{MN}\partial _{\lambda }\psi\nonumber\\&&  
-2G^{\mu i}\partial _{\mu }\psi \partial _{i}\psi 
-2G^{\mu M}\partial _{\mu }\psi \partial _{M}\theta)\nonumber\\&& 
+2G^{ij}\partial _{i}\partial _{j}\psi 
+2G^{MN}\partial _{M}\partial _{N}\theta\nonumber\\&& 
-2G^{MN}\Gamma ^{i}_{MN}\partial _{i}\psi 
-2G^{MN}\Gamma ^{P}_{MN}\partial _{P}\theta\nonumber\\&& 
-G^{ij}\partial _{i}\psi \partial _{j}\psi 
-2G^{iM}\partial _{i}\psi \partial _{M}\theta\nonumber\\&& 
-G^{MN}\partial _{M}\theta\partial _{N}\theta\,\,.
\eea
The following expressions for the Christoffel symbols are useful 
\begin{eqnarray}
G^{MN}\Gamma ^{\lambda }_{MN}(G) & = & g^{\mu \nu }\Gamma ^{\lambda }_{\mu \nu }(g)
+g^{\lambda \alpha }[(-\frac{1}{2}h^{ij}\nabla _{\alpha }h_{ij})
+(\partial _{i}V^{i}_{\alpha }+\frac{1}{2}V^{k}_{\alpha }h^{ij}\partial _{k}h_{ij})]\label{gam1} \\
G^{MN}\Gamma ^{k}_{MN}(G) & = & (h^{ij}h^{kl}\partial _{i}h_{jl}
-\frac{1}{2}h^{ik}h^{jl}\partial _{i}h_{jl})+g^{\alpha \beta }(\nabla _{\alpha }V^{k}_{\beta }
+\frac{1}{2}V^{k}_{\beta }h^{ij}\nabla _{\alpha }h_{ij})\nonumber \\
 &  & +g^{\alpha \beta }[-\partial _{i}(V^{i}_{\alpha }V_{\beta }^{k})
-\frac{1}{2}V^{i}_{\alpha }V^{k}_{\beta }h^{jl}\partial _{i}h_{jl}]\label{gam2} 
\end{eqnarray}  

Having listed the different contributions to $L$ (and $\t{L}$) 
we are now in a position to perform the reduction of the 
string effective action $\Gamma$ (and $\t{\Gamma}$).
In comparing $L$ and $\t{L}$, we notice that
a contribution involving a given number of factors of $g^{\mu \nu}$ in $L$
must be equal to the contribution from $\t{L}$ involving the same number
of factors of $\t{g}^{\mu \nu}$.
This is due to $g^{\mu \nu}=\t{g}^{\mu \nu}$ under Poisson-Lie T-duality.
Moreover, when closely examining the relations 
(\ref{omegalambda}, \ref{omegatlambdat}, \ref{lambdalambdatilde})
we realize that the Poisson-Lie duality transformations when acting on a term 
involving $n$ factors of $V^i_{\mu}$ and $m$ factors of $B'_{\mu i}$
yield a sum of terms involving $n'$ factors of $V^i_{\mu}$ 
and $m'$ factors of $B'_{\mu i}$ such that $n+m=n'+m'$.
These two facts enable us to separate the Lagrangian $L$
into different parts, each with a given number of factors of $g^{\mu \nu}$
and a given order $n+m$ in the gauge fields $V_{\mu}^i$ and $B'_{\mu i}$. 
Each part, in this separation
process must be invariant under Poisson-Lie T-duality, independently of the other parts.
This decomposition of $L$ is carried out explicitly in the appendix.

Here is an illustrative example concerning the dilatonic contribution to $L$. 
Recall that there is a Poisson-Lie T-duality invariant component in the dilaton,
namely $\psi_0(x)=\t{\psi}_0(x)$.
Therefore, equations (\ref{psi_prop},\ref{psit_prop}) imply that
$\partial_{\mu}\psi=\partial_{\mu}\t{\psi}$.
Hence, the sum of terms involving $\partial_{\mu}\psi$ in $L$ 
must be equal to the contribution in $\t{L}$ involving $\partial_{\mu}\t{\psi}$. 
In $L$, this sum is given by
\be
2G^{\mu \nu }\partial _{\mu }\partial _{\nu }\psi 
-G^{\mu \nu }\partial _{\mu }\psi \partial _{\nu }\psi\nonumber\\ 
+4G^{\mu i}\partial _{i}\partial _{\mu }\psi 
-2G^{MN}\Gamma ^{\lambda }_{MN}\partial _{\lambda }\psi\nonumber\\  
-2G^{\mu i}\partial _{\mu }\psi \partial _{i}\psi 
-2G^{\mu M}\partial _{\mu }\psi \partial _{M}\theta
\label{dmupsi_terms}
\ee
A similar expression, which is obvious to write down, holds for $\t{L}$.
When comparing the two expressions,
the first two terms are invariant under Poisson-Lie duality
due to the invariance of $G^{\mu \nu}=g^{\mu \nu}$ and $\partial_{\mu} \psi$.
As $\psi(x,y)=\psi_0(x)-\ln \det e(y)$, the  
third term identically vanishes in $L$ and $\t{L}$.
Using the Kaluza-Klein decompositions, the three remaining terms give
\be
-2\partial_{\mu}\psi 
\{g^{\lambda \sigma} \Gamma^{\mu}_{\lambda \sigma}(g)
+g^{\mu \lambda}[\partial_i V^i_{\lambda}-V^i_{\lambda} \partial_i \psi]\}\,\,.
\ee
The first term of this last expression is invariant due to
the invariance of $g^{\mu \nu}$. 
Furthermore, we have 
\be
V^k_{\mu}=[t_{\mu a}\Pi^{ab}-{u'}^b_{\mu}]{(e^{-1})}^k_b\,\,.
\ee
This last equation together with (\ref{psi_prop}) yield
\be
\partial_i V^i_{\lambda}-V^i_{\lambda} \partial_i \psi
=
f^b_{bc}(b(g)^{ca}t_{\lambda a}-{u'}_{\lambda}^c)
-\t{f}_b^{bc}a(g)_c^{\;\;\;a} t_{\lambda a}\,\,,
\label{anomalie_dilaton}
\ee
where we have used 
\bea
\partial_i \Pi^{ab}&=&e_i^c[f^a_{cd}\Pi^{bd}-f^b_{cd}\Pi^{ad}+\t{f}_c^{ab}]\,\,,
\nonumber \\
f^a_{bc} \Pi^{bc}&=&\t{f}_b^{ba}-\t{f}_b^{bc}a(g)_c^{\;\;\;a}+f^b_{bc}b(g)^{ca}\,\,.
\label{Pi_deriv} 
\eea
These relations can be found from the definition (\ref{def_pi}) of $\Pi^{ab}$ 
and by using the properties (\ref{bilinear_product2}) of the bilinear product $<,>$ 
as shown in the appendix of \cite{Sfetsos:1998pi}.
It is clear that similar relations hold for the dual Lie algebra $\t{\cal G}$.

To summarize, the comparison of the expression (\ref{dmupsi_terms}) 
and its corresponding counterpart comming from $\t{L}$ amounts
to comparing (\ref{anomalie_dilaton}) to its dual
\bea
\partial^i \t{V}_{\lambda i}-\t{V}_{\lambda i} \partial^i \t{\psi}
=
\t{f}_b^{bc}(\t{b}(\t{g})_{ca}\t{t}_{\lambda}^a-{\t{u}'}_{\lambda c})
-f^b_{bc}\t{a}(\t{g})^c_{\;\;a} \t{t}_{\lambda}^a\,\,,
\eea 
Using the Poisson-Lie duality relations, this last expression transforms into
\bea
\partial^i \t{V}_{\lambda i}-\t{V}_{\lambda i} \partial^i \t{\psi}
=
\t{f}_b^{bc}(t_{\lambda c}-\t{b}(\t{g})_{ca}{u'}_{\lambda}^a)
+f^b_{bc}\t{a}(\t{g})^c_{\;\;a} {u'}_{\lambda}^a\,\,,
\label{anomalie_duale}
\eea
The two expression (\ref{anomalie_dilaton}) and (\ref{anomalie_duale}) 
can be made equal if we impose that $f^a_{ab}=0$ and $\t{f}_a^{ab}=0$.
This ensures the vanishing
of these ``anomalous'' terms in $L$ and $\t{L}$ and guaranties, therefore, 
the invariance under Poisson-Lie duality of the dilatonic part that we are examining. 
This simple criterion, as shown in appendix B, will be sufficent to eliminate all the other 
anomalous terms comming from $L$ and $\t{L}$.
Moreover, this is consistent with the results of \cite{Tyurin:1996bu}, 
obtained by means of path integral considerations.
This is also in agreement with the non-Abelian T-duality case \cite{Bossard:2001xq}.

The comparison of the other parts of $L$ and $\t{L}$ is treated in the appendix.
The general steps in dealing with each part can be descrided as follows:
First, we search for cancellations between the terms in $R(G)$ and the dilatonic
contribution.
Then, we eliminate the derivatives of the vielbeins and the $\Pi$ matrices
using, respectively, the relations (\ref{fee}) and (\ref{Pi_deriv}).
Secondly, we eliminate as often as possible, the contributions 
involving $\Pi$ matrices. 
This is carried out by means of the relation \cite{Sfetsos:1998pi}
\be
\Omega^{abc}+{\bf c.p.}=0\,\,,
\label{Omega}
\ee
where 
\be
\Omega^{abc}=\t{f}_d^{ab}\Pi^{cd}-f^c_{ed}\Pi^{ea}\Pi^{db}\,\,.
\ee
Finally, the remaining terms are gathered in
groups of expressions wich are invariant under Poisson-Lie T-duality transformations.
Two crucial points must be mentioned here:
The invariant terms are only functions of the $x^{\mu}$ coordinates. 
while the dependence on $y^i$ is esclusively contained in the anomalous terms
(the terms not invariant under Poisson-Lie duality).
The latter vanish if we require that the structure 
constants of the two Lie algebras be traceless.  
Hence, after dropping the volume factors $\int d^n y \det e$ and  $\int d^n y \det \t{e}$,
we obtain the same expression for the original and dual
string effective actions.   

\section{ Weyl Anomaly Coefficients}

Having established the invariance of the string effective action under
Poisson-Lie duality transformations, we turn our attention now 
to the Weyl anomaly coefficients. 
At the one loop level and in the cases of Abelian \cite{Haagensen} 
and non-Abelian \cite{Bossard:2001xq} dualities,   
it has been shown that the following functional relation 
exists between the one loop Weyl anomaly coefficients 
of the original and dual sigma models:
\be
\bb^{(\t{\omega})}=
\sum_{\omega} \frac{\delta \t{\omega}}{\delta \omega}\bb^{(\omega)}\,\,,
\label{Haag_relation}
\ee
where $\omega$ designates the original backgrounds $(G_{MN},B_{MN},\Ph)$
and $\t{\omega}$ stands for the dual backgrounds $(\t{G}_{MN},\t{B}_{MN},\Pht)$.
In the case of Poisson-Lie duality , however, 
the equivalent of the above relation is not yet known.
Our aim here is to show 
that the previous relation holds for any non-linear sigma model
admitting Poisson-Lie duality.

The results of the previous section
can be recast as follows
\begin{equation}
\t{\cal L}[\t{G},\t{B},\Pht]=\chi {\cal L}[G,B,\Ph]\;\;\;\;\;{\rm where}\,\,\,
{\cal L}=\sqrt{G}e^{-\Ph}L\,\,\,,
\,\,\,\t{\cal L}=\sqrt{\t{G}}e^{-\Pht}\t{L}
\,\,,\,\,\,\chi=\frac{\det \t{e}}{\det e}\,\,.
\label{prop}
\ee
The Weyl anomaly coefficients are related to the string effective 
action $\Gamma=\int d^dx d^ny {\cal L}$ by 
\be
\bb^{(\omega)}=\sum_{\omega'}
M_{\omega \omega'}
\frac{\delta \Gamma}{\delta \omega'}
\label{bbm1}
\ee
where the matrix $M$ takes the form 
\be
M_{\omega \omega'}=-\frac{1}{\sqrt{G}e^{-\Ph}}
\left(\begin{array}{lll}
\frac{1}{2}(G_{MP}G_{NQ}+G_{MQ}G_{NP})\,\,&0\,\,&\frac{1}{2}G_{PQ}\\
0&\frac{1}{2}(G_{MP}G_{NQ}-G_{MQ}G_{NP})&0\\
\frac{1}{2}G_{MN}&0&\frac{1}{8}(D-2)
\end{array}\right)\,\, .
\ee
Note that $M$ is an invertible matrix.
Of course the same relations hold for the dual background
$\t{\omega}$ and we have
\be
\bb^{(\t{\omega})}=\sum_{\t{\omega}'}
M_{\t{\omega} \t{\omega}'}
\frac{\delta \t{\Gamma}}{\delta \tilde{\omega}'}\,\,.	
\label{bbm2}
\ee
Using the first equality of (\ref{prop}) together with the chain rule
we get
\be
\bar{\beta}^{(\t{\omega})}=
\chi \sum_{\omega'} \sum_{\t{\omega}'}
M_{\t{\omega} \t{\omega}'}
\frac{\delta \omega '}{\delta \t{\omega}'}\frac{\delta \Gamma}
{\delta \omega '}\,\,.
\label{deduced}
\ee
A crucial point in the following derivations is the invertibility
of the matrix $\frac{\delta \omega '}{\delta \t{\omega}'}$,
namely that the matrix $\frac{\delta \t{\omega}'}{\delta \omega '}$ exists.
By explicitly calculating  $\frac{\delta \t{\omega}'}{\delta \omega '}$,
along the lines of \cite{Bossard:2001xq} in the case of non-Abelian
duality, we have checked that this is indeed the case.
Using (\ref{bbm1}) into (\ref{Haag_relation}) yields
\bea
\bar{\beta}^{(\t{\omega})}=\sum_{\omega} \sum_{\omega'}
\frac{\delta \t{\omega}}{\delta \omega}
M_{\omega \omega'}\frac{\delta \Gamma}{\delta \omega'}
\eea
Comparing this last expression with (\ref{deduced})
we obtain
\be
\chi M_{\t{\omega} \t{\omega}'}=
\sum_{\omega} \sum_{\omega'}
\frac{\delta \t{\omega }}{\delta \omega}M_{\omega \omega'}
\frac{\delta \t{\omega }'}{\delta \omega'}
\label{M=M}
\ee
Therefore, showing that the Weyl anomaly coefficients are related
by $\bb^{(\t{\omega})}=
\sum_{\omega} \frac{\delta \t{\omega}}{\delta \omega}\bb^{(\omega)}$
amounts to showing that equation (\ref{M=M}) holds.

An explicit computation as in ref.\cite{Bossard:2001xq}
leads to the following relations
\bea
 M_{\lambda \lambda'} =
\det e \sum_{\omega} \sum_{\omega'}
\frac{\delta \lambda}{\delta \omega}
M_{\omega \omega'}
\frac{\delta \lambda'}{\delta \omega'}\,\,,
\label{one}
\eea
where $\lambda$ denotes 
the background $((G_0)_{MN},(B_0)_{MN},\Ph_0)$.
Similarly, we have
\bea
M_{\t{\lambda} \t{\lambda}'} =
\sum_{\lambda} \sum_{\lambda'}
\frac{\delta \t{\lambda}}{\delta \lambda}
M_{\lambda \lambda'}
\frac{\delta \t{\lambda}'}{\delta \lambda'}\,\,.
\label{two}
\eea
where $\t{\lambda}$ denotes 
the backgrounds $((\t{G}_0)_{MN},(\t{B}_0)_{MN},\Pht_0)$.
Substituting, in this last equation, $M_{\lambda \lambda'}$
as in (\ref{one}) and using the chain rule, we obtain
\bea
M_{\t{\lambda} \t{\lambda}'} =
\det e \sum_{\omega} \sum_{\omega'}
\frac{\delta \t{\lambda}}{\delta \omega}
M_{\omega \omega'}
\frac{\delta \t{\lambda}'}{\delta \omega'}\,\,.
\label{two_and_half}
\eea
Furthermore, explicit computations lead to
\bea
\det \t{e}\,\, M_{\t{\omega} \t{\omega}'} =
\sum_{\t{\lambda}} \sum_{\t{\lambda'}}
\frac{\delta \t{\omega}}{\delta \t{\lambda}}
M_{\t{\lambda} \t{\lambda}'}
\frac{\delta \t{\omega}'}{\delta \t{\lambda}'}\,\,,
\label{three}
\eea 
Plugging equation (\ref{two_and_half}) into this last expression 
and using chain rules, we obtain (\ref{M=M}). 
This concludes the proof regarding the proportionality
between the Weyl anomaly coefficients of the original sigma
model and those of its Poisson-Lie dual.

\section{ Examples }

In this section, we illustrate our analysis
by two explicit examples.
The first one concerns 
the original non-linear sigma model as given by the action 
\be
S=\int d^2 \sigma \{ f(t) \partial t
\bar{\partial} t + E_{ab}(t,g) (g^{-1} \partial g)^a
(g^{-1} \bar{\partial} g)^b -\frac{1}{4} R^{(2)} \varphi(t)
\}
\label{modele_sigma_original}
\ee
where $g$ is a group element corresponding to a 
three dimensional Lie algebra. This is taken to be of type Bianchi II.

The dual Lie algebra can be also chosen to be of type Bianchi II,
as shown in \cite{Jafarizadeh:1999xv}.
The dual sigma model is of the same form as above and is described by
\be
\t{S}=\int d^2 \sigma \{ f(t) \partial t
\bar{\partial} t + \t{E}^{ab}(t,\t{g}) (\t{g}^{-1} \partial \t{g})_a
(\t{g}^{-1} \bar{\partial} \t{g})_b -\frac{1}{4} R^{(2)} \t{\varphi}(t)
\}\,\,.
\label{modele_sigma_dual}
\ee
The non-vanishing commutation relations for the Lie algebras 
of the double are given by
\be
[T_2,T_3]=T_1\,\,,\;\;\;[\t{T}^1,\t{T}^2]=\t{T}^3\,\,,  
\ee
\be
[T_2,\t{T}^1]=-\t{T}^3\,\,,\;\;[T_3,\t{T}^2]=-T_1\,\,,\;\;
[T_3,\t{T}^1]=T_2+\t{T}^2\,\,.
\ee
Choosing a parametrization such that $g=e^{xT_1}e^{yT_2}e^{zT_3}$
for the elements of the first group and 
$\t{g}=e^{x\t{T}^1}e^{y\t{T}^2}e^{z\t{T}^3}$ 
for the elements of the second group leads to the following
matrices for $\Pi$ and  $\tilde{\Pi}$ 
\bea
\Pi^{ab}=\left(
\begin{array}{ccc}
0 & -z & 0 \\
z &  0 & 0 \\
0 &  0 & 0
\end{array}
\right)\,\,,\;\;\;
\tilde{\Pi}_{ab}=\left(
\begin{array}{ccc}
0 & 0 & 0  \\
0 & 0 & -x \\
0 & x & 0
\end{array}
\right).
\eea
Similarly, the corresponding vielbeins are given by
\bea
e^a_{\;\;\;i}=\left(
\begin{array}{ccc}
1 & z & 0 \\
0 & 1 & 0 \\
0 &  0 & 1
\end{array}
\right)\,\,,\;\;\;
\tilde{e}_a^{\;\;\;i}=\left(
\begin{array}{ccc}
1 & 0 & 0  \\
0 & 1 & 0 \\
y & 0 & 1
\end{array}
\right).
\eea
For simplicity, we choose 
$Q^{ab}=h(t) \delta^{ab}$.
This allows us to calculate the matrices $E_{ab}=(Q+\Pi)^{-1}$
and $\t{E}^{ab}=(\t{Q}+\t{\Pi})^{-1}$.

The original backgrounds and their duals
are then computed using the relations of section 2.
The metric in the original theory is found to be
\bea
G_{MN}= \left(
\begin{array}{cc}
f(t) & 0  \\
 0   & G_{ij}
\end{array}
\right)\,\,, 
\eea
where 
\bea
G_{ij}= \frac{1}{h \Delta}\left(
\begin{array}{ccc}
h^2 & zh^2 & 0 \\
zh^2 & (z^2+1)h^2 & 0\\
0 & 0  & \Delta 
\end{array}
\right)\,\,,  
\eea
Similarly, the antisymmetric tensor in the original sigma model is
\bea
B_{MN}= \frac{z}{\Delta}\left(
\begin{array}{cccc}
0 & 0 & 0 & 0 \\
0 & 0 & -1 & 0 \\
0 & 1 & 0 & 0\\
0 & 0 & 0  & 0 
\end{array}
\right)\,\,, 
\eea
with $\Delta=h^2+z^2$.
Notice that $b_{\mu \nu}$ is equal to zero in this example.
Finally, the original dilaton is 
\be
\Ph=\psi_0(t)+\frac{1}{2}\ln (h/\Delta^2)\,\,.
\ee
On the other hand, the metric in the
dual sigma model is given by
\bea
\t{G}_{MN}= \left(
\begin{array}{cc}
f(t) & 0  \\
 0   & \t{G}^{ij}
\end{array}
\right)\,\,, 
\eea
where
\bea
\t{G}^{ij}= \frac{h}{\t{\Delta}}\left(
\begin{array}{ccc}
\t{\Delta}+y^2 & 0 & y \\
0 & 1 & 0\\
y & 0  & 1 
\end{array}
\right)\,\,,  
\eea
with $\t{\Delta}=1+x^2 h^2$.
The dual antisymmetric tensor is
\bea
\t{B}_{MN}= \frac{xh^2}{\t{\Delta}}\left(
\begin{array}{cccc}
0 & 0 & 0 & 0\\
0 & 0 & -y & 0 \\
0 & y & 0 & 1\\
0 & 0 & -1 & 0 
\end{array}
\right)\,\,. 
\eea
The dual dilaton is written as
\be
\Pht=\psi_0(t)+\frac{1}{2}\ln (h^3 / \t{\Delta}^2)\,\,.
\ee
We come now to the string effective actions corresponding to these
two sigma models.
Notice that in this example the determinants of the vielbeins are both 
equal to one.
This implies an equality between the integration factors $\sqrt{G}e^{-\Ph}$
and $\sqrt{\t{G}}e^{-\Pht}$. They are both equal to $\sqrt{f(t)} e^{-\psi_0(t)}$.
Finally, the two Lagrangians $L[G,B,\Ph]$ and $\t{L}[\t{G},\t{B},\Pht]$, 
yield the same expression
\be
L=\t{L}=-\frac{1}{2} \left[
h+\frac{1}{h}+\frac{3}{2}f^{-1} {\left( \frac{\dot{h}}{h} \right)}^2
+f^{-1}(2 {\dot{\psi}_0}^2 -4 \ddot{\psi}_0)
\right]\,\,.
\ee
Here, a dot stands for differentiation with respect to $t$.

One realizes that this expression is invariant under
the transformation $h \rightarrow 1/h$ with $f$ and $\psi_0$ kept unchanged.
The expressions for the two Lagrangians
do not depend on the $y^i$ coordinates as expected
from the general conclusions of section 3.
Furthermore, there are no anomalous parts as the structure constants of  
Bianchi II Lie algebras are traceless $f^a_{ab}=\t{f}_a^{ab}=0$. 
It is worth mentioning that we have carried out similar calculations 
based on this double but with more complicated
choices for the tensor $Q^{ab}$. In each case, the conclusions
reached in section 3 were confirmed.

The second example is chosen to illustrate how the presence 
of ``anomalous'' terms, 
resulting from the non-vanishing traces of the structure
constants, breaks the proportionality between the original and 
dual string effective actions.
For this purpose, we take the two sigma models,
the original and its dual, to have
the same form as in the first example 
(\ref{modele_sigma_original},\ref{modele_sigma_dual}).
The only difference is that now the two Lie algebras of the double 
are Borel type of dimension two.
A representation of this double (second paper of \cite{Klimcik}) is
\bea
T_1= \left(
\begin{array}{cc}
1 & 0  \\
0 & 0 \\
\end{array}
\right)\,\,,\;\;\; 
T_2= \left(
\begin{array}{cc}
0 & 1  \\
0 & 0 \\
\end{array}
\right)\,\,,\;\;\;
\t{T}^1= \left(
\begin{array}{cc}
0 & 0  \\
0 & 1 \\
\end{array}
\right)\,\,,\;\;\;
\t{T}^2= \left(
\begin{array}{cc}
0  & 0  \\
-1 & 0 \\
\end{array}
\right)\,\,.
\eea
The group elements
are parametrized by $g=e^{x T_1}e^{y T_2}$ and $\t{g}=e^{x \t{T}^1}e^{y \t{T}^2}$.
The matrices $\Pi$ and $\t{\Pi}$ are as follows
\bea
\Pi^{ab}= \left(
\begin{array}{cc}
 0 & y \\
-y & 0 \\
\end{array}
\right)\,\,,\;\;\; 
\t{\Pi}_{ab}= \left(
\begin{array}{cc}
0 & y  \\
-y & 0 \\
\end{array}
\right)\,\,,
\eea
and the vielbeins are 
\bea
e^a_{\;\;\;i}= \left(
\begin{array}{cc}
1 & 0 \\
y & 1 \\
\end{array}
\right)\,\,,\;\;\; 
\t{e}_a^{\;\;\;i}= \left(
\begin{array}{cc}
1 & 0 \\
y & 1 \\
\end{array}
\right)\,\,.
\eea
We choose also $Q^{ab}=h(t)\delta^{ab}$.
The original backgrounds are listed below
\bea
G_{MN}= \left(
\begin{array}{cc}
f(t) & 0  \\
 0   & G_{ij}
\end{array}
\right)\,\,, 
\eea
where  
\bea
G_{ij}=\frac{h}{\Delta'}\left( 
\begin{array}{cc}
1+y^2 & y \\
y & 1 
\end{array}
\right)\,\,,
\eea
and 
\bea
B_{MN}=\frac{y}{\Delta'}\left( 
\begin{array}{ccc}
0 & 0 & 0 \\
0 & 0 & -1 \\
0 & 1 & 0 
\end{array}
\right)\,\,,
\nonumber \\
\Ph=\psi_0(t)+ \ln (h/\Delta')\,\,,
\eea
where $\Delta'={h(t)}^2+y^2$.
The dual backgrounds are obtained from these by simply replacing $h$
by $1/h$. This is due to the equality between the matrices $\Pi$ and 
$\t{\Pi}$ and the vielbeins $e$ and $\t{e}$.
As in the first example, the determinants of the vielbeins are equal to one. 
Hence, the integration factors in the two string effective 
actions are such that
$\sqrt{G}e^{-\Ph}=\sqrt{\t{G}}e^{-\Pht}=\sqrt{f}e^{-\psi_0}$.
The string effective Lagrangian corresponding to the original sigma model
is found to be
\be
L=-\left[2(h+\frac{1}{h})+\frac{1}{2}f^{-1}\left(\frac{\dot{h}}{h}\right)^2
+f^{-1}({\dot{\psi}_0}^2-2\ddot{\psi}_0) + \frac{8y^2}{h}\right]\,\,.
\ee
The dual string effective Lagrangian $\t{L}$ is found by replacing 
$h$ by $1/h$ in this last expression.
The last term, $ -8y^2 /h$, is not invariant under this transformation
and therefore the two Lagrangians $L$ and $\t{L}$ are not equal.
This anomaly is due to the non-vanishing traces of the structure constants
of the Borel double. 
In fact, this anomalous term can be recovered from the sum
\bea
{(S_0^{-1})}_{ad}[-f^a_{bc} \Pi^{bc} f^d_{ef} \Pi^{ef}
+2f^a_{bc} \Pi^{bc} \t{f}^{ed}_e
+2f^e_{bc} \Pi^{bc} f^a_{ef} \Pi^{fd}  
-2f^e_{eb} \t{f}^{ba}_c \Pi^{cd}
-2\t{f}^{be}_e f^a_{cb} \Pi^{cd}
]\,\,.
\label{anomaly0}
\eea 
It can be shown, using (\ref{Pi_deriv}), that this combination involves either
$f^a_{ab}$ or $\t{f}_a^{ab}$ which do not vanish for the Borel double. 

\section{Conclusion}

In this paper, we have analyzed the quantum equivalence
of sigma models related by Poisson-Lie T-duality transformations.
Our results are obtained at the level of the one loop string effective action.  
An appropriate reparametrization \`a la Kaluza-Klein of the various string backgrounds
has been used to give a simpler form to the Poisson-Lie duality transformations.
As a consequence, it has been possible to cast the Lagrangian of the string effective action
into independently Poisson-Lie duality invariant parts. On the other hand, the non-invariant terms
have been shown to vanish if we impose the conditions $f^a_{ab}=0$ and $\t{f}_a^{ab}=0$
on the structure constants of the two Lie algebras of the Drinfeld double.
These conditions are the same as those suggested in \cite{Tyurin:1996bu} in a path integral derivation
of Poisson-Lie duality. Furthermore, we deduce a functional relation between the Weyl anomaly coefficients
corresponding to two non-linear sigma model related by Poisson-Lie duality. In particular, 
a conformally invariant sigma model leads, under Poisson-Lie duality, 
to a dual theory with the same property.

When dealing with the anomalous terms resulting from the reduction of 
the string effective action one wonders whether the sufficient conditions
that we have imposed (the vanishing of the traces of structure constant) are also necessary conditions.
This is all the more a natural question since in \cite{Balog} two sigma models related by Poisson-Lie
duality have been shown to possess the same one loop beta functions. 
The equivalence holds in spite of 
a non-vanishing trace for one of the structure constants corresponding to one 
of the two Lie algebras forming the 
Drinfeld double. This is clearly in contradiction with our conclusions.  
The authors of ref.\cite{Balog}  
use, however, a dilaton transformation which differs from ours. 
Their analysis is carried out in a strict field theory
sense, regardless of the relationship between sigma models and string theory effective actions.  
The dilaton shift is precisely related to the diffeomorphism transformation that allowed 
them to conclude the equivalence of the 
two beta functions.  The results of \cite{Balog} 
can be interpreted in two different ways:
Either it is a coincidence, due to the particularity of their model, that led to the possibility
of absorbing the anomalous terms into their dilaton shift 
(which is the same as a diffeomorphism transformation), or this 
is symptomatic of a more general phenomenon. The problem of Poisson-Lie duality 
when some of the structure constants have 
non-vanishing traces, obviously, requires further investigation. For instance, 
it would be desirable to compute more
physical quantities like the free energy in order to check the real equivalence, 
under Poisson-Lie duality, of 
two non-linear sigma models. Finally, we strongly believe that the dilaton 
transformation, given in here in the 
context of string theory, is the correct one. These are the only transformations 
which lead to a proportionality
between the integration weights $\sqrt{G}e^{-\Ph}$ and $\sqrt{\t{G}}e^{-\Pht}$. 
This is an essential requirement in demanding the invariance 
under Poisson-Lie duality of the string effective action.   
The dilaton transformation under any T-duality is, nevertheless,
a complex issue as shown in \cite{dilaton}.

Another interesting question is the investigation of the invariance of the string effective
action under a much larger group of dualities, where Poisson-Lie duality might be just a special case.
In other words, one would like to find the equivalent of the  $O(d,d)$ transformation group
present in the case of string backgrounds having Abelian symmetries \cite{Maharana,Meissner}. 
This is suggested by our observation that the transformations of some intermediate backgrounds,
introduced in order to simplify Poisson-Lie T-duality transformations,  are precisely of the 
form of Abelian T-duality.  Due to the complexity of the general expression 
of the string effective Lagrangian $L$, however, we are not able to make 
this possible hidden symmetry manifest.    
 
A natural extension of our work would be to push the analysis beyond the 
leading order in loop expansion,
as has been done for Abelian duality \cite{kaloper1,Parsons}.
The resemblence of Poisson-Lie duality and Abelian duality, as mentioned above, 
makes us think that a similar treatment along the lines of 
\cite{kaloper1,Parsons} is possible.
It is worth mentioning that there are two ways of dealing with duality 
transformations in general: The first consists 
in correcting the duality transformations order by order in the string 
perturbation parameter $\alpha'$ \cite{kaloper1,Parsons}.
In the second approach, however, one keeps the duality transformations as 
given by the one loop order and deforms, instead,
the non-linear sigma model \cite{Bonneau:2001za}. 
The equivalence of these two methods needs a closer examination.

Finally, our results could be of interest in cosmology models based 
on the string effective action. This is 
in the spirit of the ideas presented in \cite{pbb}. 
The so-called pre-big-bang scenario takes,
as a starting point, a string effective action where 
the backgrounds possess Abelian isometries.
However, many of the interesting models in cosmology 
are based on metrics invariant under the action of 
non-Abelian isometries.  The contruction of their 
duals under Poisson-Lie duality is therefore possible. 
This is one of the motivations of a recent work \cite{Jafarizadeh:1999xv} 
where all the Drinfeld doubles based on Bianchi type Lie algebras are classified.
We should mention that string cosmology requires the introduction of a 
Lagrangian for matter fields, which must be invariant under duality 
transformations. It has been shown in \cite{Nurmagambetov:2001iy}
that this is indeed possible if one chooses fundamental strings as 
gravitational sources.  
The role of Poisson-Lie duality in string cosmological 
is currently under investigation.

\vskip 2 cm
\noindent
{\bf Aknowledgments:}
We would like to thank Janos Balog, Peter Forg\'acs and Ian Jack
for useful discussions.
\appendix

\section{Useful Identities}
Before dealing with the string effective Lagrangian $L$, 
we would like to list some of the relations that are essential 
in obtaining the reduced string effective action.
First, we find it useful to introduce, 
in addition to the notations of the main text,
the following definitions
\bea
t_{\mu \nu a}
&=&
\partial_{\mu} t_{\nu a}-\partial_{\nu} t_{\mu a}\,\, ,
\\
{u'}^a_{\mu \nu}
&=&
\partial_{\mu} {u'}^a_{\nu}-\partial_{\nu} {u'}^a_{\mu}\,\, ,
\\
{u^{(1)}}_{\mu \nu}^b
&=&
{u'}^b_{\mu \nu}+t_{\mu \nu a}(v_0)^{ab} \,\,,
\\
{h'}_{\mu \nu}^c
&=&
{u'}^a_{\mu}{u'}^b_{\nu}f^c_{ab}
+(t_{\mu a}{u'}^b_{\nu}-t_{\nu a}{u'}^b_{\mu})\t{f}_b^{ac} \,\, ,
\\
{F'}_{\mu \nu d}
&=&
t_{\mu a}t_{\nu b}\t{f}_d^{ab}
-(t_{\mu a}{u'}^b_{\nu}-t_{\nu a}{u'}^b_{\mu})f^a_{bd} \,\,,
\\
{h''}_{\mu \nu}^c
&=&
{h'}_{\mu \nu}^c - {F'}_{\mu \nu d} (v_0)^{dc} \,\,,
\\
\alpha_{\mu \nu a}
&=&
t_{\mu \nu a}-{F'}_{\mu \nu a} \,\,,
\\
{\gamma'}^a_{\mu \nu}
&=&
{u'}^a_{\mu \nu}+{h'}^a_{\mu \nu}\,\,.
\eea
The relations that we have employed are 
\bea
{(e^{-1})}^m_a \partial_m {(e^{-1})}^i_b-{(e^{-1})}^m_b 
\partial_m {(e^{-1})}^i_a
&=&
f^c_{ab} {(e^{-1})}^i_c \,\, , 
\\
V^k_{\mu}(x,y)
&=&
[t_{\mu a}(x)\Pi^{ab}(y)-{u'}^b_{\mu}(x)]{(e^{-1})}^k_b(y) \,\, ,
\\
V^k_{\mu \nu}(x,y)
&=&
[t_{\mu \nu a}(x)\Pi^{ab}(y)-{u'}^b_{\mu \nu}(x)]{(e^{-1})}^k_b(y)\,\,,
\\
{B'}_{\mu k}(x,y)
&=&
-t_{\mu a}(x)e^a_k(y) \,\, ,
\\
{B'}_{\mu \nu k}(x,y)
&=&
-t_{\mu \nu a}(x)e^a_k(y) \,\, ,
\\
{B'}_{i \mu j}(x,y)
&=&
t_{\mu a} f^a_{bc} e^b_i e^c_j  
\\
&=&
[t_{\mu \nu a}(A_0)^{ab}-{u^{(1)}}_{\mu \nu}^b]{(e^{-1})}^k_b(y)\,\,, 
\\
F_{\mu \nu}^i
&=&
[{F'}_{\mu \nu d}(A_0)^{dc}+{h''}_{\mu \nu}^c]{(e^{-1})}^k_b(y)\,\,, 
\\
h^{(1)}_{\mu \nu i}
&=&
-[t_{\mu \nu a}(S_0)^{ab}S_{bc}+-{u^{(1)}}_{\mu \nu}^a A_{ac}]e^c_i \,\,,
\\
h^{(2)}_{\mu \nu i}
&=&
[{F'}_{\mu \nu a}(S_0)^{ab}S_{bc}-{h''}_{\mu \nu}^a A_{ac}]e^c_i\,\,. 
\eea

\section{Reduction of the String Lagrangian}
In this appendix, we will collect the terms in the original 
string effective Lagrangian $L=R-\frac{1}{12}H^2+2\nabla^2 \Ph-(\nabla \Ph)^2$.
These are grouped according to the number of factors of $g^{\mu \nu}$.
This separation is vital as $g^{\mu \nu}$ is invariant under
Poisson-Lie T-duality ($g^{\mu \nu}=\t{g}^{\mu \nu}$).
Therefore, each of these terms must be independently invariant.
The expression of the dual string Lagrangian $\t{L}$
is obtained from $L$ by changing tilded quantities
into untilded ones and vice versa.
The Poisson-Lie duality transformations (\ref{lambdalambdatilde}) 
are then used to show the equality between a given term in $L$, 
involving a given number of factors of $g^{\mu \nu}$, 
and its counterpart in $\t{L}$.
 
\subsection{Order zero in $g^{\mu \nu}$}
The terms without any factors of $g^{\mu \nu}$ give
\bea
&&-h^{ij}\partial_i \psi \partial_j \psi +2\partial_i(h^{ij}\partial_j \psi)
-\partial_i \partial_j h^{ij}
+\frac{1}{2}h^{km} h^{il} h^{jn} (\partial_k h_{ij} \partial_l h_{mn}-
\partial_k b_{ij} \partial_l b_{mn})
\nonumber \\
&&-\frac{1}{4}h^{kl} h^{im} h^{jn}(\partial_k h_{ij} \partial_l h_{mn}
+\partial_k b_{ij} \partial_l b_{mn})\,\,.
\eea
After substituting for $h_{ij}$, $b_{ij}$ and $\psi$ together with the
use of equations (\ref{Omega}) to handle some of the terms
involving $\Pi$, we find two main expressions.
The first, invariant under Poisson-Lie T-duality, is given by 
\bea
&&-\{f^c_{ad} \t{f}_i^{jk}{(\t{S}_0^{-1})}^{di}{(S_0^{-1})}_{ck}{v_0}^{ab}{(S_0^{-1})}_{bj}\}
\nonumber \\
&&+\frac{1}{2} \{f^i_{aj} f^k_{cl} {(\t{S}_0^{-1})}^{jl} {v_0}^{ab}{(S_0^{-1})}_{bk}
{v_0}^{cd}{(S_0^{-1})}_{di}  
+ \t{f}_i^{aj}\t{f}_k^{cl} {(S_0^{-1})}_{jl} {(\t{v}_0)}_{ab}{(\t{S}_0^{-1})}^{bk}
{(\t{v}_0)}_{cd}{(\t{S}_0^{-1})}^{di}\}  
\nonumber \\
&&-\frac{1}{4} \{\t{f}_a^{cd} \t{f}_b^{ij} {(\t{S}_0^{-1})}^{ab} {(S_0^{-1})}_{ci}
{(S_0^{-1})}_{dj}
+ f^m_{ak} f^n_{ci} {(S_0^{-1})}_{mn} {(\t{S}_0^{-1})}^{ac} {(\t{S}_0^{-1})}^{ik}\}  
\nonumber \\
&&-\{f^a_{ab} f^c_{cd} {(\t{S}_0^{-1})}^{bd} +\t{f}_a^{ab} \t{f}_c^{cd} {(S_0^{-1})}_{bd}\}
-\frac{1}{2}\{f^a_{bc}f^b_{ad}{(\t{S}_0^{-1})}^{cd}
+\t{f}_a^{bc}\t{f}_b^{ad}{(S_0^{-1})}_{cd}\}
\nonumber \\
&&+\frac{1}{2} \{f^k_{ci} \t{f}_a^{lm} {v_0}^{ab}{(S_0^{-1})}_{bk} {v_0}^{cd}{(S_0^{-1})}_{dl}
{v_0}^{ij}{(S_0^{-1})}_{jm}\}
\nonumber \\
&&+\frac{3}{2} \{f^d_{mn} \t{f}_c^{mn} {v_0}^{ci}{(S_0^{-1})}_{id}\}
-2 \{f^m_{mc} \t{f}_n^{nd} {v_0}^{ci}{(S_0^{-1})}_{id}\}
-\{f^m_{nc} \t{f}_m^{nd} {v_0}^{ci}{(S_0^{-1})}_{id}\}\,\,.   
\eea
In this expression (and in the rest of this appendix), 
each contribution between curly brackets
is equal to its dual counterpart.
The second expression is an
anomalous part (not invariant under Poisson-Lie duality)
and can be cast in the form 
\bea
&&[2f^a_{bc} f^b_{di} \Pi^{di} {v_0}^{cj} {(S_0^{-1})}_{ja}
+2 f^b_{bc} f^a_{di} \Pi^{di} {v_0}^{cj} {(S_0^{-1})}_{ja} 
+f^a_{ab} f^b_{cd} \Pi^{di} {v_0}^{cj} {(S_0^{-1})}_{ij}]
\nonumber \\
&&+{(S_0^{-1})}_{mn}[-f^m_{bd} \Pi^{bd} f^n_{ij} \Pi^{ij}
+2 f^m_{bd} \Pi^{bd} \t{f}_c^{cn}
+2 f^b_{di} \Pi^{di} f^m_{bc} \Pi^{cn}
-2 f^d_{db} \t{f}_c^{bm} \Pi^{cn}
\nonumber \\
&&-2\t{f}_d^{db} f^m_{cb} \Pi^{cn}]\,\,. 
\eea 
It is clear that this contribution vanishes if we impose
$f^a_{ab}=\t{f}_a^{ab}=0$.

\subsection{Order one in $g^{\mu \nu}$}
There are three independently invariant contributions 
at this order in $g^{\mu \nu}$.
The first one does not involve the gauge fields $V_{\mu}^i$ 
and $B'_{\mu i}$.
The second contains one gauge field (either $V_{\mu}^i$ 
or $B'_{\mu i}$).
The third depends on a combination of two gauge fields.

\subsubsection{order zero in gauge fields}
This order is equal to
\be
-\frac{1}{4}g^{\mu\nu}h^{im}h^{jn}
(\partial_{\mu} h_{ij} \partial_{\nu} h_{mn}
+\partial_{\mu} b_{ij} \partial_{\nu} b_{mn})
\label{zero1}
\ee
The following relations are needed for the reduction of this term  
\bea
\partial S_{ab} &=& 
-\partial (S_0)^{cd}[S_{ac}S_{db}+A_{ac}A_{db}]
-\partial (A_0)^{cd}[S_{ac}A_{db}+A_{ac}S_{db}] 
\\
\partial A_{ab} &=& 
-\partial (A_0)^{cd}[S_{ac}S_{db}+A_{ac}A_{db}]
-\partial (S_0)^{cd}[S_{ac}A_{db}+A_{ac}S_{db}]
\eea
where $\partial$ stands for either $\partial_{\mu}$ or $\partial_i$.
This is a consequence of $(S+A)=(S_0+A_0)^{-1}$ 
combined with the properties
\bea
A_0(x,y)=v_0(x)+\Pi(y)\,\,\,\Rightarrow\,\,\,
\partial_{\mu} A_0=\partial_{\mu} v_0
\eea
After substituting for $h_{ij}$ and $b_{ij}$,
this zeroth order yields 
\be
-\frac{1}{4}g^{\mu\nu}{(S_0^{-1})}_{ac}{(S_0^{-1})}_{bd}
[\partial_{\mu}(S_0)^{ab} \partial_{\nu}(S_0)^{cd}
+\partial_{\mu}(v_0)^{ab} \partial_{\nu}(v_0)^{cd}]\,\,.
\label{n1}
\ee
We deduce that the equivalent expression 
coming from the dual Lagrangian is 
\be
-\frac{1}{4}g^{\mu\nu}{(\t{S}_0^{-1})}^{ac}{(\t{S}_0^{-1})}^{bd}
[\partial_{\mu}(\t{S}_0)_{ab} \partial_{\nu}(\t{S}_0)_{cd}
+\partial_{\mu}(\t{v}_0)_{ab} \partial_{\nu}(\t{v}_0)_{cd}]\,\,.
\label{n2}
\ee
It can be shown that equations (\ref{n1}) and (\ref{n2}) are identical
upon using 
\bea
\partial (S_0)^{ab} &=&  
-\partial (\t{S}_0)_{cd}[(S_0)^{ac}(S_0)^{db}+(v_0)^{ac}(v_0)^{db}]
\nonumber \\
&&-\partial (\t{v}_0)_{cd}[(S_0)^{ac}(v_0)^{db}+(v_0)^{ac}(S_0)^{db}]\,\,, 
\nonumber \\
\partial (v_0)^{ab} &=& 
-\partial (\t{v}_0)_{cd}[(S_0)^{ac}(S_0)^{db}+(v_0)^{ac}(v_0)^{db}]
\nonumber \\
&&-\partial (\t{S}_0)_{cd}[(S_0)^{ac}(v_0)^{db}+(v_0)^{ac}(S_0)^{db}]\,\,.
\eea
These are obtained from $(S_0+v_0)=(\t{S}_0+\t{v}_0)^{-1}$.

\subsubsection{order one in gauge fields}
We find two expressions at this order.
The first is
\be
2g^{\mu \nu} \nabla_{\mu} [\partial_i V^i_{\nu}-V^i_{\nu} \partial_i \psi]
=
2g^{\mu \nu} \nabla_{\mu} 
[f^b_{bc}(b(g)^{ca}t_{\nu a}-{u'}_{\nu}^c)
-\t{f}_b^{bc}a(g)_c^{\;\;\;a} t_{\nu a}]\,\,.
\ee
This is an anomalous contribution which vanishes when
$f^a_{ab}=\t{f}_a^{ab}=0$.
The remaining contribution to this order is found to be 
\bea
&&g^{\mu \nu}
[\frac{1}{2}h^{im}h^{jn}(\partial_{\nu}h_{ij}V^k_{\mu}\partial_k h_{mn}
+\partial_{\mu} b_{ij} V^k_{\nu} \partial_k b_{mn}) 
+ h^{ik}\partial_{\mu}h_{kj}\partial_i V^j_{\nu}
\nonumber \\
&&+h^{im}h^{jn}\partial_{\mu}b_{ij}
(\partial_m {B'}_{\nu n}+\partial_m V^k_{\nu} b_{kn})]
\eea
In the reduction of this last expression, the terms
proportional to $\Pi$ and $\Pi^2$ have been eliminated 
through the use of $\Omega^{abc}+{\bf c.p.}=0$.
The final expression reduces to
\bea
&&g^{\mu \nu}\Big\{
\frac{1}{2}{u'}^e_{\mu}\t{f}^{ad}_e {(S_0^{-1})}_{ab}
\partial_{\nu}(v_0)^{bc}{(S_0^{-1})}_{cd} \nonumber \\
&&-\frac{1}{2}t_{\mu a}f^a_{cb}
[ -2\partial_{\nu} {S_0}^{cd}{(S_0^{-1})}_{de}{v_0}^{eb} 
+\partial_{\nu}{v_0}^{cb}
+{v_0}^{cd}{(S_0^{-1})}_{de}\partial_{\nu}{v_0}^{ef}
{(S_0^{-1})}_{fg}{v_0}^{gb} 
]\Big\}
\nonumber \\
&&+
g^{\mu \nu}\Big\{
-t_{\mu a} \t{f}^{ab}_c 
[\partial_{\nu} {S_0}^{cd}{(S_0^{-1})}_{db}
-{v_0}^{cd}{(S_0^{-1})}_{de}\partial_{\nu}{v_0}^{ef}{(S_0^{-1})}_{fb}]
\nonumber \\
&&+{u'}^e_{\mu}f^a_{ef}
[{(S_0^{-1})}_{ab}\partial_{\nu}{v_0}^{bc}{(S_0^{-1})}_{cd}{v_0}^{df}
-{(S_0^{-1})}_{ab}\partial_{\nu}{S_0}^{bf}]
\Big\} 
\eea 
Here again each expression between curly brackets is equal 
to its dual counterpart.

\subsubsection{order two in gauge fields}
At this order, the first contribution is an anomalous one and is given by
\bea
-2g^{\mu \nu} \partial_i 
[V^i_{\mu}(\partial_j V^j_{\nu}-V^j_{\nu}\partial_j \psi)]
+g^{\mu \nu}(\partial_i V^i_{\mu}+V^i_{\mu}\partial_i \psi)
(\partial_j V^j_{\nu}-V^j_{\nu}\partial_j \psi)\,\,.
\eea
Both of these terms contain
$\partial_j V^j_{\nu}-V^j_{\nu}\partial_j \psi$
which has been already encountered.
This part vanishes when $f^a_{ab}=\t{f}_a^{ab}=0$.
The second contribution is the invariant part coming from
\bea
-\frac{1}{4}g^{\mu \nu}h^{im}h^{jn}[F_{\mu i j}F_{\nu m n}
+{h^{(1)}}_{\mu i j}{h^{(1)}}_{\nu m n}]
\eea
where we have introduced the quantities
\bea
F_{\mu i j}
&=&
V^k_{\mu} \partial_k h_{ij}
+\partial_i V^k_{\mu} h_{kj}
+\partial_j V^k_{\mu} h_{ki}\,\,, 
\nonumber \\
h^{(1)}_{\mu i j}
&=&
V^k_{\mu} \partial_k b_{ij}
+[\partial_i {B'}_{\mu j}+ (\partial_i V^k_{\mu})b_{kj}
-(i \leftrightarrow j)]\,\,.
\eea
It can then be shown that the sum of these terms is equal to
the following 
\bea
&&-\frac{1}{4}g^{\mu \nu}\Big[
\{{u'}^n_{\mu}\t{f}_n^{ab}{u'}^i_{\nu}\t{f}_i^{cd}
{(S_0^{-1})}_{ac}{(S_0^{-1})}_{bd}
+t_{\mu e}f^e_{ab}t_{\nu f}f^f_{cd}{(\t{S}_0^{-1})}^{ac}{(\t{S}_0^{-1})}^{bd}\}
\nonumber \\
&&+2\{\lambda_{\mu m}^a \lambda_{\nu i}^c {(\t{S}_0^{-1})}^{mi} {(S_0^{-1})}_{ac}\}
+2\{\lambda_{\mu m}^a \lambda_{\nu i}^d {(S_0^{-1})}_{db}{v_0}^{bm}
{(S_0^{-1})}_{ac}{v_0}^{ci}\}
\nonumber \\
&&-2\{{u'}_{\mu}^i \t{f}^{jk}_i t_{\nu n} f^n_{ab} {(S_0^{-1})}_{kd}{v_0}^{db}
{(S_0^{-1})}_{jc}{v_0}^{ca}\}
\nonumber \\
&&+4\{t_{\mu p}f^p_{mn} \lambda^c_{\nu i} {(\t{S}_0^{-1})}^{in}
{(S_0^{-1})}_{cd}{v_0}^{dm}
-{u'}^p_{\mu} \t{f}_p^{ab} \lambda^c_{\nu i}{(S_0^{-1})}_{ac}
{(S_0^{-1})}_{bd}{v_0}^{di}\} 
\nonumber \\
&&+2\{ {u'}^c_{\mu} {u'}^d_{\nu} f^b_{ac} f^a_{bd}
+t_{\mu c} t_{\nu d} \t{f}^{cb}_a \t{f}^{da}_b\}
+2 \{t_{\mu d} {u'}^c_{\nu} \t{f}_c^{ab} f^d_{ab}\}
+4 \{t_{\mu d} {u'}^c_{\nu} \t{f}_b^{ad} f^b_{ac}\}
\Big]
\eea
In this last expression 
$\lambda^a_{\mu c}=t_{\mu b} \t{f}^{ba}_c+{u'}^b_{\mu} f_{bc}^a$,
and has the following property under Poisson-Lie duality:
$\t{\lambda}^a_{\mu c}=-\lambda^a_{\mu c}$.
We have checked that each quantity between curly brackets is invariant.

\subsection{Order two in $g^{\mu \nu}$}

Here we can treat at the same time orders two, three and four 
in the gauge fields.
The contribution to $L$, at this order, is
\bea
&&-\frac{1}{4}g^{\mu \rho}g^{\nu \lambda} \Big[ 
h_{ij}(V_{\mu \nu}^i-F_{\mu \nu}^i)
(V_{\rho \lambda}^j-F_{\rho \lambda}^j)
+h^{ij}(h^{(1)}_{\mu \nu i}+h^{(2)}_{\mu \nu i})
(h^{(1)}_{\rho \lambda j}+h^{(2)}_{\rho \lambda j})
\Big]\,\,.
\eea
Using the relations in appendix A, one can show that this
contribution reduces to
\bea
&&-\frac{1}{4}g^{\mu \rho}g^{\nu \lambda}\Big[
\{\alpha_{\mu \nu a}\alpha_{\rho \lambda b}
[(S_0)^{ab}-(A_0)^{ac}(S_0^{-1})_{cd}(A_0)^{db}]
+{\gamma '}_{\mu \nu}^a {\gamma '}_{\rho \lambda}^b (S_0^{-1})_{ab}\} 
\nonumber \\
&&+\{\alpha_{\mu \nu a}(v_0)^{ab} (S_0^{-1})_{bc}{\gamma '}_{\rho \lambda}^c
-{\gamma '}_{\mu \nu}^a (S_0^{-1})_{ab} (v_0)^{bc} \alpha_{\rho \lambda c}\} 
\Big]\,\,.
\eea
Each expression between curly brackets is invariant
under Poisson-Lie duality.
This is demonstrated using the duality transformations
\bea
\t{\alpha}=-\gamma'\;\;\;,\;\;\;
\t{\gamma'}=-\alpha\,\,,
\nonumber \\
S_0-A_0 S_0^{-1} A_0=\t{S}_0^{-1}\;\;\;,\;\;\;
v_0 S_0^{-1}=-\t{S}_0^{-1} \t{v}_0\,\,.
\eea
   
\subsection{Order three in $g^{\mu \nu}$: invariance of $h_{\mu \nu \rho}$ }

This contribution comes from the term 
$h_{\mu \nu \rho}h^{\mu \nu \rho}$ in the expression of $H_{MNP}H^{MNP}$
in (\ref{hh}).
We will show here that $h_{\mu \nu \rho }=\t{h}_{\mu \nu \rho }$.
Recall that
\be
h_{\mu \nu \rho }=(\partial _{\rho }b_{\mu \nu })
+(-V_{\rho }^{k}\partial _{k}b_{\mu \nu })
+\frac{1}{2}(V^{k}_{\mu \rho }B_{\nu k}'+B_{\mu \rho k}'V_{\nu }^{k}) 
+\frac{1}{2}(F^{k}_{\rho \mu }B_{\nu k}'+V_{\rho }^{k}V_{\mu }^{l}B_{l\nu k}')
+{\bf c.p.}
\label{trois}
\ee  
Under Poisson-Lie duality $b_{\mu \nu }=\t{b}_{\mu \nu }$.
Moreover, both $b_{\mu \nu }$ and $\t{b}_{\mu \nu }$
are independent of the coordinates $y^i$.
Therefore, the first term is invariant and the second vanishes.
The third term which, upon using using the relations
in appendix A, gives
\be
\frac{1}{2}(V^{k}_{\mu \rho}B_{\nu k}'+B_{\mu \rho k}'V_{\nu }^{k})+{\bf c.p.}
=
\frac{1}{2}(t_{\mu \rho a}{u'}_{\nu }^a+{u'}^a_{\mu \rho }t_{\nu a})+{\bf c.p.}
\ee
The duality transformations $t_{\mu a}=-{\t{u}'}_{\mu a}$
and ${u'}^a_{\mu}= -{\t{t}}^a_{\mu}$ 
clearly show that this combination is invariant.
Finally, the fourth term reduces to 
\bea
\frac{1}{2}(F^{k}_{\rho \mu }B_{\nu k}'+V_{\rho }^{k}V_{\mu }^{l}B_{l\nu k}')+{\bf c.p.}
&=&\Big[\{(-{u'}^a_{\mu}{u'}^b_{\nu}t_{\rho c}f^c_{ab}
+t_{\mu a}t_{\nu b}{u'}^c_{\rho}\t{f}_c^{ab}\} \nonumber \\
&& -t_{\mu a} t_{\nu b} t_{\rho c} \Omega^{abc}\Big] + {\bf c.p.} 
\eea
The first two terms are invariant under Poisson-Lie duality
while the third one vanishes due to  
$\Omega^{abc}+\Omega^{bca}+\Omega^{cab}=0$.  



\end{document}